\documentclass[prb,twocolumn,aps,showpacs,eqsecnum,amsmath]{revtex4}
\usepackage{graphicx}
\usepackage{bm}

\begin{document}
\title{
Generalized two-leg Hubbard ladder at half-filling:\\
Phase diagram and quantum criticalities
}
\author{M.\ Tsuchiizu and A.\ Furusaki}
\affiliation{
Yukawa Institute for Theoretical Physics, Kyoto University,
Kyoto 606-8502, Japan}
\date{\today}

\begin{abstract}
The ground-state phase diagram of the half-filled two-leg Hubbard
ladder with inter-site Coulomb repulsions and exchange coupling
is studied by using the strong-coupling perturbation theory and the 
weak-coupling bosonization method.
Considered here as possible ground states of the ladder model are
four types of density-wave states with different angular momentum
($s$-density-wave state, $p$-density-wave state,
$d$-density-wave state, and $f$-density-wave state)
and four types of quantum disordered states, i.e., Mott insulating
states (S-Mott, D-Mott, S'-Mott, and 
D'-Mott states, where S and D stand for $s$- and $d$-wave symmetry).
The $s$-density-wave state, the $d$-density-wave state, and
the D-Mott state are also known as the
charge-density-wave state, the staggered-flux state, and
the rung-singlet state, respectively.
Strong-coupling approach naturally leads to the Ising model in a
transverse field as an effective theory for the quantum
phase transitions between the staggered-flux state and the
D-Mott state and between the charge-density-wave state and
the S-Mott state,
where the Ising ordered states correspond to doubly degenerate ground
states in the staggered-flux or the charge-density-wave state.
From the weak-coupling bosonization approach it is shown that
there are three cases in the quantum phase transitions between a
density-wave state and a Mott state: the Ising (Z$_2$) criticality,
the SU(2)$_2$ criticality, and a first-order transition.
The quantum phase transitions between Mott states and between
density-wave states are found to be the U(1) Gaussian criticality.
The ground-state phase diagram is determined by integrating
perturbative renormalization-group equations.
It is shown that the S-Mott state and the staggered-flux state
exist in the region sandwiched by the charge-density-wave phase
and the D-Mott phase.
The $p$-density-wave state, the S'-Mott state, and the D'-Mott state
also appear in the phase diagram when
the next-nearest-neighbor repulsion is included.
The correspondence between 
Mott states in extended Hubbard ladders 
and spin liquid states in spin ladders
is also discussed.
\end{abstract}


\maketitle

\section{Introduction}

Ladder systems have been studied intensively over the years
as a simplified model system that shows
variety of quantum phenomena due to strong electron
correlations.\cite{Dagotto}
Since the ladder models can be analyzed with powerful nonperturbative
methods such as bosonization and conformal field theory
as well as with large-scale numerical calculations,
they provide a useful testing ground of various
theoretical ideas developed for the two-dimensional case.
Moreover, the studies of ladder systems have been strongly
stimulated by experimental developments in synthesizing
compounds with ladder structure that show superconductivity
and spin-liquid behavior.\cite{Azuma,Ishida,Kojima}
A good example is the ladder compound
Sr$_{14}$Cu$_{24}$O$_{41}$ that shows $d$-wave superconducting
order\cite{Uehara} under pressure with Ca doping and
charge-density-wave (CDW) order as recently suggested
experimentally.\cite{Blumberg,Gorshunov} 
Theoretical studies on doped ladder models such as the Hubbard
and $t$-$J$ ladders
\cite{Dagotto1992,Finkelstein,Fabrizio,Noack,Sigrist,Tsunetsugu1994,%
      Khveshchenko1994,Nagaosa,Dagotto,Gopalan,Sano,%
      Schulz1996,Balents1996,Orignac,Yoshioka1997,Tsuchiizu2001}
have established that the dominant correlation is indeed
a $d$-wave-like superconducting order, a feature that is
reminiscent of the $d$-wave superconductivity in high-$T_c$ cuprates.
On the other hand, undoped half-filled Hubbard and 
  Heisenberg ladders are insulators that
have a gap in both charge and spin
excitations.\cite{Dagotto,Gopalan,Khveshchenko1994,Shelton,Lin,%
       Noack,White1994,LeHur}
This spin-liquid behavior is caused by singlet formation on each
rung, and the state is said to be in the rung-singlet phase.
It is also named D-Mott phase\cite{Lin} because of its close
connection to the $d$-wave-like paring state.

Recent theoretical interest on the ladder models has been focused on
the search of exotic phases in these systems.
In particular, the staggered-flux (SF) state,\cite{Marston}
which is also known as the orbital
antiferromagnet\cite{Halperin,Nersesyan1989,Schulz1989} and
the $d$-density wave,\cite{Nayak2000,Chakravarty}
has received
a lot of attention.\cite{Nersesyan1991,Nersesyan1993,Ivanov1998,%
Scalapino2001,Tsutsui2001,Fjaerestad}
For more than a decade the SF state has been intensively studied in
connection
with the pseudo-gap phase in the two-dimensional high-$T_c$
cuprates.\cite{Marston,Sachdev2000,Lee,Ivanov2000,Leung,Nayak2000,%
               Chakravarty,Nayak2002}
The SF state has spontaneous currents flowing around
plaquettes, breaking the time-reversal symmetry.
Even though ladders are one-dimensional (1D), the long-range order
of the SF correlation is possible at half-filling, since the
symmetry broken in this state is discrete.
This point was emphasized recently in Ref.\ \onlinecite{Fjaerestad},
where it is also suggested that the SF phase should occur in the phase
diagram of the SO(5) symmetric Hubbard
model.\cite{Scalapino1998,Frahm}
Besides the SF phase, the ground-state phase diagram of the ladder
models can include the D-Mott phase mentioned above,
the CDW phase,\cite{Vojta1999} and other phases.

Motivated by these developments, in this paper we attempt 
systematic exploration of the ground-state phase diagram of
a generalized two-leg Hubbard ladder at half-filling
that has not only repulsive on-site and inter-site
interactions but also antiferromagnetic (AF) exchange interaction and
pair hoppings between the legs.
To map out the possible phases in the parameter space of the model
and to analyze various quantum phase transitions,
we employ both the strong-coupling perturbation theory and
the weak-coupling bosonization method.
We find that the inclusion of the additional interactions leads to
emergence of various new phases.

In the strong-coupling approach, we describe the SF state
as an AF ordered state of pseudo-spins
that represent currents flowing on the rungs.
The effective theory near the phase boundary between the SF state
and the D-Mott state is then found to be the 1D Ising
model in a transverse field.
The D-Mott phase is thus interpreted as a disordered state of
the Ising model.
We also present a similar mapping to the 1D quantum Ising model
for the quantum phase transition between the CDW phase and the
S-Mott phase.\cite{Lin}
Here the CDW state and the S-Mott state correspond to the ordered and
quantum disordered states of the Ising model, respectively.
Furthermore, we show that a low-energy effective theory near the
phase transition between the D-Mott and the S-Mott phases is the
XXZ spin chain in a staggered field, which exhibits a U(1)
Gaussian criticality.

In the weak-coupling limit,
we follow the standard approach of taking continuum limit and
bosonizing the Hamiltonian.
We obtain a coupled sine-Gordon model for four bosonic modes
(charge/spin \& even/odd modes) and analyze it by perturbative
renormalization-group (RG) method and a semiclassical approximation.
The scaling equations we derive are equivalent to those obtained
earlier by Lin, Balents, and Fisher.\cite{Lin}
We depart here from the earlier work.
We consider four types of density-wave states with different
angular momentum:\cite{Nayak2000} $s$-density wave (= CDW),
$p$-density wave (PDW, which is equivalent to the spin-Peierls state),
$d$-density wave (= SF), and $f$-density wave (FDW).
These density-wave states break Z$_2$ symmetry and can have
long-range order at zero temperature.
We find that in general there should appear four types of Mott
insulating phases (called S-Mott, D-Mott, S'-Mott, and D'-Mott
states), each of which can be obtained as a quantum
disordered state from one of the four Z$_2$-symmetry-breaking
density-wave states.
We then study quantum phase transitions among these 8 phases
and show that a transition between a density-wave state and a
Mott state is either second order (in the Ising or SU(2)$_2$
universality class) or first order.\cite{transitions}
Phase transitions between density-wave states and between Mott
states are U(1) Gaussian criticalities.
After classifying the phases and the quantum phase transitions,
we determine the ground-state phase diagram of the extended
Hubbard model with extra inter-site repulsion and the exchange
interaction.
We find that the S-Mott and the SF phases appear in the
parameter space of couplings where the D-Mott and the CDW phases
compete.
We also show that the next-nearest-neighbor repulsion stabilizes
the S'-Mott state and the PDW state; the latter state is connected to
the D-Mott state through the SU(2)$_2$ criticality.

This paper is organized as follows.
In Sec.\ \ref{sec:model}
   the model we analyze in this paper is introduced.
In Sec.\ \ref{sec:strongcoupling}
   we study the ground-state phase diagram by
   the strong-coupling perturbation theory, and
   examine phase transitions between the competing 
   ground states: the SF, D-Mott, CDW, and S-Mott states.
In Sec.\ \ref{sec:weakcoupling} we apply the weak-coupling
bosonization method to study the ground-state phase diagram.
We derive effective low-energy theory for the charge mode and
for the spin mode that describe the Gaussian, Ising, and SU(2)$_2$
criticalities.
The connection of our results to the phase diagram of spin ladders
with spin liquid ground states is also discussed.
We then determine the phase diagram of the generalized Hubbard
ladder from perturbative RG equations.
Finally, the results are summarized in Sec.\ \ref{sec:summary}.

\section{Model}\label{sec:model}

We consider a half-filled two-leg Hubbard ladder
   with on-site and inter-site Coulomb repulsions
   and rung exchange interaction.
The Hamiltonian we study in this paper is given by
\begin{equation}
H
= 
H_{t_\parallel}
+ 
H_{t_\perp}
+
H_{\mathrm{int}}
+
H_{V_\parallel}
+
H_{V'}
+
H_{\mathrm{pair}}.
\label{eq:H}
\end{equation}
The first two terms
   describe hopping along and between the legs, respectively:
\begin{eqnarray}
H_{t_\parallel} &=& - t_\parallel \sum_{j,\sigma,l}
   (c_{j,l,\sigma}^\dagger \, c_{j+1,l,\sigma}^{}+\mathrm{H.c.})
,
\\
H_{t_\perp} &=& - t_\perp \sum_{j,\sigma}
   (c_{j,1,\sigma}^\dagger \, c_{2,j,\sigma}^{}+\mathrm{H.c.})
,
\end{eqnarray}
where $c_{j,l,\sigma}$ annihilates an
   electron of spin $\sigma(=\uparrow,\downarrow)$ on rung $j$ and
   leg $l(=1,2)$.
The Hamiltonian
$H_{\mathrm{int}}=H_U+H_{V_\perp}+H_{J_\perp}$ consists of three
terms representing interactions within a rung: the on-site repulsion,
\begin{equation}
H_{U}
=
U \sum_{j,l} n_{j,l,\uparrow} \, n_{j,l,\downarrow}
,
\end{equation}
the nearest-neighbor repulsion on a rung,
\begin{equation}
H_{V_\perp}
=
 V_\perp \sum_j n_{j,1} \, n_{j,2}
,
\end{equation}
and the nearest-neighbor exchange interaction on a rung,
\begin{equation}
H_{J_\perp}
=
 J_\perp \sum_{j} \bm{S}_{j,1} \cdot \bm{S}_{j,2} .
\end{equation}
The density operators are 
   $n_{j,l,\sigma}=c_{j,l,\sigma}^\dagger \, c_{j,l,\sigma}^{}$
   and $n_{j,l}=n_{j,l,\uparrow}+n_{j,l,\downarrow}$, and
   the spin-$\frac{1}{2}$ operator is given by
\begin{equation}
\mbox{\boldmath $S$}_{j,l} = \frac{1}{2} \sum_{\sigma_1,\sigma_2}
   c_{j,l,\sigma_1}^\dagger \, \bm{\sigma}_{\sigma_1,\sigma_2} \,
   c_{j,l,\sigma_2}^{},
\end{equation}
   where $\bm{\sigma}_{\sigma_1,\sigma_2}$ are the Pauli matrices.
The Hamiltonian (\ref{eq:H}) also has nearest-neighbor repulsive
interaction within a leg,
\begin{equation}
H_{V_\parallel}
=
 V_\parallel \sum_{j,l} n_{j,l} \, n_{j+1,l}
,
\label{eq:HVpara}
\end{equation}
and next-nearest-neighbor repulsion,
\begin{equation}
H_{V'}
=
 V' \sum_{j} 
  \left(n_{j,1} \, n_{j+1,2} + n_{j,2} \, n_{j+1,1} \right)
.
\label{eq:HVprime}
\end{equation}
The last component of the Hamiltonian (\ref{eq:H}) is
   the pair hopping between the legs,
\begin{equation}
H_{\mathrm{pair}}
=
t_\mathrm{pair} \sum_j 
\left(
  c_{j,1,\uparrow}^\dagger \, c_{j,1,\downarrow}^\dagger \, 
  c_{j,2,\downarrow}^{} \, c_{j,2,\uparrow}^{} 
+ \mathrm{H.c.}
\right)
.
\label{eq:Hpair}
\end{equation}
The coupling constants, $U$, $V_\perp$, $V_\parallel$, $V'$,
$J_\perp$, and $t_{\mathrm{pair}}$,
   are assumed to be either zero or positive.
(Most of our discussions are actually concerned with the case
$V_\parallel=V'=t_\mathrm{pair}=0$.)
In this paper we consider only the half-filled case where
$\sum_{j,l}n_{j,l}$ equals the number of total lattice sites.

\section{Strong-Coupling Approach}\label{sec:strongcoupling}

In this section, we perform strong-coupling analysis starting from the
independent rungs and discuss transitions between various
insulating phases.

We begin with eigenstates of $H_{\mathrm{int}}$ for decoupled
rungs at half-filling.
Convenient basis states for two electrons on a single rung (e.g.,
$j$th rung) with $S^z_{j,1}+S^z_{j,2}=0$ are
\begin{eqnarray}
|1\rangle_j &=&
\left| 
  \begin{array}{c}
     \uparrow \\ \downarrow
  \end{array}
\right\rangle_j
\equiv 
c_{j,1,\uparrow}^\dagger \, c_{j,2,\downarrow}^\dagger \, | 0 \rangle 
, 
\label{eq:state1}
\\
|2\rangle_j &=&
\left| 
  \begin{array}{c}
     \downarrow \\ \uparrow
  \end{array}
\right\rangle_j
\equiv 
c_{j,1,\downarrow}^\dagger \, c_{j,2,\uparrow}^\dagger \, | 0 \rangle
,
\\
|3\rangle_j &=&
\left| 
  \begin{array}{c}
     \uparrow\downarrow \\ - 
  \end{array}
\right\rangle_j
\equiv 
c_{j,1,\uparrow}^\dagger \, c_{j,1,\downarrow}^\dagger \, | 0 \rangle 
, 
\\
|4\rangle_j &=&
\left| 
  \begin{array}{c}
     - \\ \uparrow\downarrow
  \end{array}
\right\rangle_j
\equiv 
c_{j,2,\uparrow}^\dagger \, c_{j,2,\downarrow}^\dagger \, | 0 \rangle 
.
\label{eq:state4}
\end{eqnarray}
The interaction Hamiltonian $H_{\mathrm{int}}$ is diagonalized as
\begin{eqnarray}
H_{\mathrm{int}} \frac{| 1 \rangle_j - | 2 \rangle_j }{\sqrt{2}}
&=& 
\left(V_\perp - \frac{3}{4}J_\perp \right)
 \frac{| 1 \rangle_j - | 2 \rangle_j }{\sqrt{2}}
,
\label{eq:eigen1}
\\
H_{\mathrm{int}} \frac{| 1 \rangle_j + | 2 \rangle_j }{\sqrt{2}}
&=& 
\left(V_\perp + \frac{1}{4}J_\perp \right)
 \frac{| 1 \rangle_j + | 2 \rangle_j }{\sqrt{2}}
,
\\
H_{\mathrm{int}} \, | 3 \rangle_j
&=& 
U \, |3 \rangle_j
,
\\
H_{\mathrm{int}} \, | 4 \rangle_j
&=& 
U \, |4 \rangle_j
.
\label{eq:eigen4}
\end{eqnarray}

Comparing the eigenvalues, we find that
   the lowest-energy state of $H_{\mathrm{int}}$ for
   $U>V_\perp -3J_\perp/4$ is
\begin{eqnarray}
|\mbox{D-Mott}\rangle
= 
\prod_j
\frac{1}{\sqrt{2}} 
\left[ \left| 
  \begin{array}{c}
     \uparrow \\ \downarrow
  \end{array}
\right\rangle_j 
- 
\left| 
  \begin{array}{c}
     \downarrow \\ \uparrow
  \end{array}
\right\rangle_j
\right]
.
\label{eq:D-Mott}
\end{eqnarray}
This state is a direct product of rung singlets and is nothing
but the strong-coupling limit of the D-Mott phase\cite{Lin} or the Mott
insulating phase of a half-filled Hubbard ladder.

When $U<V_\perp -3J_\perp/4$, on the other hand, the doubly occupied
   states $|3\rangle$ and $|4\rangle$ become lowest-energy states.
In this case, one of the possible ground states
    is the on-site paired insulating state realized in the S-Mott
   phase,\cite{Lin}
\begin{equation}
|\mbox{S-Mott}\rangle
= 
\prod_j
\frac{1}{\sqrt{2}} 
\left[
\left| 
  \begin{array}{c}
     \uparrow\downarrow \\ - 
  \end{array}
\right\rangle_j
+
\left| 
  \begin{array}{c}
     - \\ \uparrow\downarrow
  \end{array}
\right\rangle_j
\right]
.
\label{eq:S-Mott}
\end{equation}
Another possible ground state is the CDW state:
\begin{subequations}
\begin{equation}
|\mathrm{CDW}\rangle_1=
\prod_j 
\left[
\left| 
  \begin{array}{c}
     \uparrow\downarrow \\ - 
  \end{array}
\right\rangle_{2j-1}
\left| 
  \begin{array}{c}
     - \\ \uparrow\downarrow
  \end{array}
\right\rangle_{2j}
\right]
\label{eq:CDW1}
\end{equation}
and
\begin{equation}
|\mathrm{CDW}\rangle_2=
\prod_j 
\left[
\left| 
  \begin{array}{c}
     - \\ \uparrow\downarrow
  \end{array}
\right\rangle_{2j-1}
\left| 
  \begin{array}{c}
     \uparrow\downarrow \\ - 
  \end{array}
\right\rangle_{2j}
\right].
\label{eq:CDW2}
\end{equation}
\label{eq:CDW}
\end{subequations}

In the next subsections we study phase transitions between these
phases.

\subsection{CDW--S-Mott transition: Ising criticality}

In this subsection we discuss the phase transition
   between the S-Mott phase\cite{Lin} and the CDW
   phase\cite{Lin,Vojta1999} for $U<V_\perp-3J_\perp/4$.
This can be analyzed by mapping
   the system onto an effective spin model.
A similar analysis for the SO(5) symmetric ladder is reported
in Refs.\ \onlinecite{Scalapino1998} and \onlinecite{Frahm}.

We restrict ourselves to the lowest-energy states 
   $|3\rangle$ and $|4\rangle$ and denote them as
\begin{eqnarray}
|+\rangle_j
\equiv
\left|3\right\rangle_j 
, \,\,\,
|-\rangle_j
\equiv
  \left|4\right\rangle_j 
\label{eq:Ising_spin}
\end{eqnarray}
to make the connection to a spin model more evident.
We regard $|\pm\rangle$ as the pseudo-spin up/down states.
In this picture, the antiferromagnetic ordering of the spins
   corresponds to the CDW ordering.
We will treat the single-particle hopping terms $H_{t_\parallel}$ and
   $H_{t_\perp}$ as weak perturbations to derive effective Hamiltonian 
   in the Hilbert space of $|+\rangle$ and $|-\rangle$.
The lowest-order contributions come from the second-order processes: 
\begin{eqnarray}
H^{(2a)} 
&=& 
H_{t_\parallel} \, 
   \frac{1}{E_0-H_\mathrm{int}} \, H_{t_\parallel},
\\
H^{(2b)} 
&=& 
H_{t_\perp} \, \frac{1}{E_0-H_\mathrm{int}} \, H_{t_\perp},
\end{eqnarray}
   where $E_0=NU$ with $N$ being the number of rungs.
The nonzero matrix elements of $H^{(2a)}$ and $H^{(2b)}$
   are given by
\begin{eqnarray}
\langle \pm, \mp |
H^{(2a)} | \pm , \mp \rangle_j
&=&
\frac{4t_\parallel^2}{U-2V_\perp},
\\
\langle \pm | H^{(2b)} | \pm \rangle_j
=
\langle \pm | H^{(2b)} | \mp \rangle_j
&=&
\frac{2t_\perp^2}{U-V_\perp+3J_\perp/4}
,
\nonumber \\
\end{eqnarray}
   where
   $|s,s'\rangle_j \equiv |s\rangle_j |s'\rangle_{j+1}$ ($s,s'=\pm$). 
The above Hamiltonian is written in terms of pseudo-spin operators as
\begin{eqnarray}
H^{(2a)}
&=&
\frac{2t_\parallel^2}{2V_\perp-U} \sum_j
\left(\tau_j^z \, \tau_{j+1}^z -1\right),
\\
H^{(2b)}
&=& 
\frac{2t_\perp^2}{U-V_\perp+3J_\perp/4} 
   \sum_j \tau_j^x + \mathrm{const.}
,
\end{eqnarray}
   where $\tau_j^z$ and $\tau_j^x$ are Pauli matrices
   acting on the pseudo-spin states:
   $\tau_j^z|\pm\rangle_j = \pm |\pm\rangle_j$  and
   $\tau_j^x|\pm\rangle_j=|\mp\rangle_j$.
Here we find that $H^{(2a)}$ favors \emph{antiferromagnetic} ordering,
   while $H^{(2b)}$ prevents the order.
We thus find that the effective Hamiltonian for the doubly occupied
   states $H^\mathrm{eff}_\mathrm{CS}=H^{(2a)}+H^{(2b)}$ is given by 
   the one-dimensional quantum Ising model, 
\begin{eqnarray}
H^\mathrm{eff}_\mathrm{CS}
&=& 
\sum_j \left( K \, \tau_j^z \, \tau_{j+1}^z 
  - h \, \tau_j^x \right)
,
\label{eq:H^eff_CS}
\end{eqnarray}
   where the antiferromagnetic exchange coupling $K$ and
   the magnitude of the transverse field $h$ are given by
\begin{eqnarray}
K
= 
\frac{2t_\parallel^2}{2V_\perp-U},
\,\,\,
h = 
\frac{2t_\perp^2}{V_\perp-3J_\perp/4-U}
.
\end{eqnarray}
This model exhibits the Ising criticality at $K=h$ between
   the ordered phase (i.e., the CDW phase) 
   for $K>h$ and the disordered phase for $K<h$.
The ground state in the disordered phase is essentially
   the eigenstate of $\tau^x$ with eigenvalue $+1$, which is nothing 
   but the S-Mott phase:
\begin{eqnarray}
|\tau^x\!=\!+1\rangle_j
= 
\frac{|+\rangle_j + |-\rangle_j}{\sqrt2}
\to |\mbox{S-Mott}\rangle
.
\end{eqnarray}

The condition for the CDW phase to appear is
   given in terms of the Hubbard interactions as
\begin{eqnarray}
V_\perp
>
\frac{1-(t_\perp/t_\parallel)^2}{1-2(t_\perp/t_\parallel)^2} U
+ \frac{3}{4[1-2(t_\perp/t_\parallel)^2]} J_\perp
,
\label{eq:CDW-Smott}
\end{eqnarray}
   where $0<t_\perp/t_\parallel<1/\sqrt{2}$.
When $t_\perp/t_\parallel>1/\sqrt{2}$,
   the CDW phase is not realized within our approximation.

Here we briefly discuss effects of $H_{V_\parallel}$, $H_{V'}$, and
   $H_{\mathrm{pair}}$, treating them as small perturbations.
The lowest-order contributions come from the first-order perturbation,
   $H^{(1a)}=H_{V_\parallel}+H_{V'}$ and  $H^{(1b)}=H_{\mathrm{pair}}$,
   which can be written in terms of the pseudo-spin operators as
   $H^{(1a)}=2V_\parallel \sum_j (\tau^z_j \, \tau^z_{j+1}+1)
            -2V' \sum_j (\tau^z_j \, \tau^z_{j+1}-1)$
   and $H^{(1b)}=t_{\mathrm{pair}} \sum_j \tau^x_j$. 
The coupling constants in the quantum Ising model are modified to
\begin{eqnarray}
K
\!\! &=& \!\! 
\frac{2t_\parallel^2}{2V_\perp-U} + 2V_\parallel-2V',
\\
h
\!\! &=&\!\! 
\frac{2t_\perp^2}{V_\perp-3J_\perp/4-U}
-t_\mathrm{pair}
.
\end{eqnarray}
Thus, $H_{V_\parallel}$, $H_{V'}$, and $H_{\mathrm{pair}}$
   do not change the Ising universality and only affects the coupling
   constants.
Their main effect is to move the phase boundary.
The $V_\parallel$ and $t_{\mathrm{pair}}$ interactions
   favor the Ising ordered phase or the CDW phase, while
   the $V'$ interaction is in favor of the S-Mott phase.

\subsection{D-Mott--S-Mott transition: Gaussian criticality}

Next we discuss the parameter region $U\approx V_\perp-3J_\perp/4$.
In this case the low-energy states of $H_{\mathrm{int}}$ are formed
   out of 
   $(|1\rangle_j-|2\rangle_j)/\sqrt{2}$, $|3\rangle_j$, and
   $|4\rangle_j$; see Eqs.\ (\ref{eq:eigen1})-(\ref{eq:eigen4}).
The analysis in the previous subsection indicates that,
   among the states made of $|3\rangle_j$ and $|4\rangle_j$, only
   the S-Mott phase can appear for $U\approx V_\perp-3J_\perp/4$ due
   to the large transverse field $h$.
We thus keep only the two states,
\begin{equation}
|+\rangle\!\rangle_j\equiv\frac{|1\rangle_j-|2\rangle_j}{\sqrt{2}},
\quad
|-\rangle\!\rangle_j\equiv\frac{|3\rangle_j+|4\rangle_j}{\sqrt{2}},
\label{eq:D-Mott S-Mott}
\end{equation}
for each rung and derive an effective 
low-energy Hamiltonian for these states to study the competition
between the S-Mott and D-Mott phases.
In this basis,
$H_{\mathrm{int}}$ and $H_{t_\perp}$ on the $j$th rung read
\begin{eqnarray}
H_{\mathrm{int}}
&=& 
\left(
\begin{array}{cc}
V_\perp-\frac{3}{4}J_\perp   &  0 \\
0  &    U       
\end{array}
\right)
,
\label{eq:Hint_G}
\\
H_{t_\perp}
&=& 
\left(
\begin{array}{cc}
0          &  -2t_\perp \\
-2t_\perp  &    0       
\end{array}
\right)
,
\end{eqnarray}
   where $|+\rangle\!\rangle_j= {}^t(1 ,0)$ and
   $|-\rangle\!\rangle_j= {}^t(0 ,1)$.
Since we are interested in the region near the level crossing point 
   $U=V_\perp-3J_\perp/4$, we split the Hamiltonian as
\begin{equation}
H_{\mathrm{int}}+H_{t_\perp}+H_{t_\parallel}
 = H^{(0)}_\mathrm{DS} + H'_\mathrm{DS} ,
\end{equation}
   where the unperturbed Hamiltonian $H^{(0)}_\mathrm{DS}$ and 
   the perturbation term $H'_\mathrm{DS}$ are given by
   $H^{(0)}_\mathrm{DS}=U\sum_j(n_{j,1,\uparrow} \, n_{j,1,\downarrow}
       + n_{j,2,\uparrow} \, n_{j,2,\downarrow}
       + n_{j,1} \, n_{j,2})$ and
   $H'_\mathrm{DS}=(V_\perp-U)\sum_j n_{j,1} \, n_{j,2} + H_{J_\perp}
       + H_{t_\perp} + H_{t_\parallel}$. 
Up to second order in $H'_\mathrm{DS}$ the effective Hamiltonian is
   obtained as $H^{(0)}+H^{(1)}+H^{(2)}$:
\begin{eqnarray}
&&
H^{(0)}_j
=
\left(
\begin{array}{cc}
 U    &  0  \\
 0    &  U       
\end{array}
\right)
,
\\
&&
H^{(1)}_j
=
\left(
\begin{array}{cc}
-(U-V_\perp+\frac{3}{4}J_\perp)    &   -2t_\perp  \\
 -2t_\perp    &      0       
\end{array}
\right)
,
\label{eq:Gauss_H1}
\\
&&
H^{(2)}
=
H_{t_\parallel}
\frac{1}{E_0-H_0} 
H_{t_\parallel},
\label{eq:Gauss_H2}
\end{eqnarray}
   where $H^{(0)}=\sum_j H^{(0)}_j$, $H^{(1)}=\sum_j H^{(1)}_j$,
 and $E_0=NU$.
Now we introduce spin-1/2 operators
  $\widetilde{S}^x_j$, $\widetilde{S}^y_j$, and $\widetilde{S}^z_j$ and 
  identify the two states $|+\rangle\!\rangle_j$ and
  $|-\rangle\!\rangle_j$ with up and down states of the pseudo-spin
 $\widetilde{S}^z_j$. 
The first-order term $H^{(1)}$ (\ref{eq:Gauss_H1}) is then written as
\begin{eqnarray}
H^{(1)}
&=&
-\left( U-V_\perp+\frac{3}{4}J_\perp \right)
 \sum_j \left(\widetilde{S}^z_j+\frac{1}{2}\right)
\nonumber \\
&& {}
-4t_\perp \sum_j \widetilde{S}^x_j
.
\label{eq:H^1}
\end{eqnarray}
The energy difference between the $|\pm\rangle\!\rangle_j$ states and
   the rung hopping
   are represented as the longitudinal and
   transverse  magnetic fields, respectively.
The nonzero matrix elements of $H^{(2)}$ (\ref{eq:Gauss_H2}) are given 
   by
\begin{eqnarray}
\langle\!\langle \pm, \pm | H^{(2)} | \pm , \pm \rangle\!\rangle_j
&=&
- \frac{2t_\parallel^2}{U},
\\
\langle\!\langle \pm, \pm | H^{(2)} | \mp , \mp \rangle\!\rangle_j
&=&
+ \frac{2t_\parallel^2}{U},
\\
\langle\!\langle \pm, \mp | H^{(2)} | \pm , \mp \rangle\!\rangle_j
&=&
- \frac{t_\parallel^2}{2U},
\\
\langle\!\langle \pm, \mp | H^{(2)} | \mp , \pm \rangle\!\rangle_j
&=&
+ \frac{t_\parallel^2}{2U},
\end{eqnarray}
where
 $|s,s'\rangle\!\rangle_j\equiv
  |s\rangle\!\rangle_j|s'\rangle\!\rangle_{j+1}$ ($s,s'=\pm$).
Thus the second-order contribution $H^{(2)}$
  is written in terms of the pseudo-spin operators as
\begin{eqnarray}
H^{(2)}
&=&
-\frac{t_\parallel^2}{U}
   \sum_j \left(3 \widetilde{S}^z_j \widetilde{S}^z_{j+1}
                +\frac{5}{4}\right)
\nonumber \\
&& {}
+\frac{2t_\parallel^2}{U} 
 \sum_j \left( \widetilde{S}^+_j \widetilde{S}^+_{j+1}
               + \widetilde{S}^-_j \widetilde{S}^-_{j+1} \right)
\nonumber\\
&& {}
+\frac{t_\parallel^2}{2U} 
   \sum_j \left( \widetilde{S}^+_j \widetilde{S}^-_{j+1}
               + \widetilde{S}^-_j \widetilde{S}^+_{j+1} \right)
.
\label{eq:H^2}
\end{eqnarray}
From Eqs.\ (\ref{eq:H^1}) and (\ref{eq:H^2}) we find that,
  for $U\approx V_\perp-3J_\perp/4$, 
  the low-energy effective Hamiltonian
  $H^{\mathrm{eff}}_{\mathrm{DS}}=H^{(1)}+H^{(2)}$
  is given by the anisotropic spin chain under the longitudinal and
   transverse magnetic fields:
\begin{eqnarray}
H^{\mathrm{eff}}_{\mathrm{DS}}
&=&
\sum_j 
\left[
J^x \, \widetilde{S}^x_j \widetilde{S}^x_{j+1}
-J^{yz} \left( \widetilde{S}^y_j \widetilde{S}^y_{j+1}
   + \widetilde{S}^z_j \widetilde{S}^z_{j+1} \right)
\right]
\nonumber \\
&&{}
- \sum_j \left( h^x \widetilde{S}^x_j + h^z \widetilde{S}^z_j \right)
,
\label{eq:Heff_Gaussian}
\end{eqnarray}
   where $J^x=5t_\parallel^2/U$, $J^{yz}=3t_\parallel^2/U$, 
   $h^x = 4t_\perp$,
   and $ h^z = U-V_\perp+3J_\perp/4 $.
We are interested in the case where the Zeeman field in the $z$
   direction $h^z$ is weak.
When $h^z=0$, $H^{\mathrm{eff}}_{\mathrm{DS}}$ is equivalent to the
   XXZ model with the exchange anisotropy $\Delta=J^x/J^{yz}=5/3$ and
   a uniform field in the $z$ direction.
It is known \cite{Alcaraz,Cabra} that the XXZ model is in the massless
   phase governed by the $c=1$ conformal field theory (CFT) with a
   compactification radius $R$ ($1/2\sqrt{\pi}<R<1/\sqrt{\pi}$), if
   the uniform field is in the range
    $0.175J^{yz} \lesssim h^x < \frac{8}{3}J^{yz}$.
The weak perturbation $h^z$ is acting on this gapless system.
From the transformation
   $\widetilde{S}^{y,z}_j\to(-1)^j\widetilde{S}^{y,z}_j$ we see that
   the Zeeman field $h^z$ acts as a staggered transverse field in the
   antiferromagnetic XXZ model.
Since the scaling dimension of $(-1)^j\widetilde{S}^{y,z}$ is
   $\pi R^2$, it is a relevant perturbation leading to the opening of
   a gap.\cite{Oshikawa} 

Hence we find that, when $h^z \neq 0$,
   the $h^z$ term is always relevant and generates a mass gap,
   while for $h^z=0$ the system reduces to the $c=1$ CFT or the
   Gaussian model.
Therefore the D-Mott--S-Mott transition
   is a Gaussian U(1) criticality
   with the central charge $c=1$.
The critical point is at $h^z=0$, i.e., 
\begin{eqnarray}
U-V_\perp+ \frac{3}{4} J_\perp = 0.
\label{eq:crit-DS}
\end{eqnarray}
The character of the gapped phases at $h^z\ne0$ is deduced by looking
   at the dominant $h^z$-term.
Since the gapped phases should correspond to states minimizing the
   relevant $h^z$-term, $-h^z\sum_j\widetilde{S}^z_j$,
   in Eq.\ (\ref{eq:Heff_Gaussian}),
   we conclude that for $h^z>0$ ($h^z<0$)
   the ground state is a ferromagnetically ordered state with positive
   (negative) magnetization $\langle\widetilde{S}^z\rangle$,
   or equivalently, in the D-Mott (S-Mott) phase
   in the original Hubbard ladder model;
   see Eq.\ (\ref{eq:D-Mott S-Mott}).

\begin{figure}[t]
\includegraphics[width=6.cm]{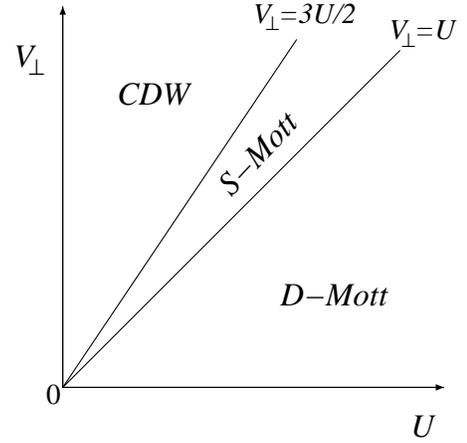}
\caption{
Strong-coupling phase diagram of
$H_{t_\parallel}+H_{t_\perp}+H_\mathrm{int}$ at
$t_\perp=t_\parallel/2$ and 
   $J_\perp=0$.
The CDW--S-Mott transition is in the Ising universality class,
   while the S-Mott--D-Mott transition is in the U(1)
   (Gaussian) universality class.
The CDW (S-Mott) phase corresponds to the ordered (disordered) phase
  in the effective quantum Ising model (\ref{eq:H^eff_CS}).
The S-Mott and D-Mott phases are the ferromagnetically ordered phases
   of the effective spin model (\ref{eq:Heff_Gaussian}).
}
\label{fig:strong1}
\end{figure}
\begin{figure}[t]
\includegraphics[width=7cm]{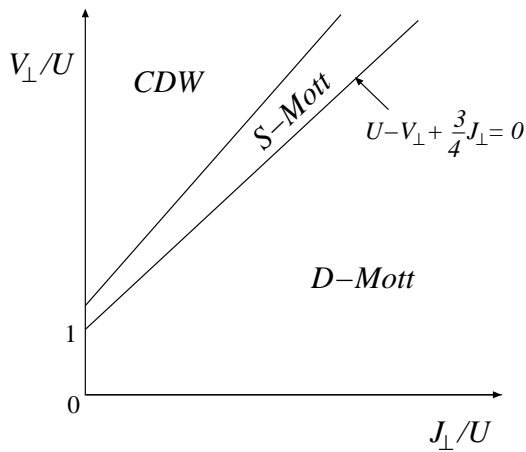}
\caption{
Strong-coupling phase diagram of
$H_{t_\parallel}+H_{t_\perp}+H_\mathrm{int}$ at $t_\perp=t_\parallel/2$
on the plane of $V_\perp/U$ and $J_\perp/U$.
The CDW phase occupies the parameter region where the condition
   (\protect{\ref{eq:CDW-Smott}}) is satisfied.
}
\label{fig:strong2}
\end{figure}

The phase diagram obtained from the strong-coupling perturbation
   theory is shown in Fig.\ \ref{fig:strong1}, where parameters are
   taken as $t_\perp=t_\parallel/2$ and $J_\perp=0$.
The phase transition between the D-Mott state and the
   S-Mott state is described as the Gaussian criticality,
   while the phase transition between the S-Mott state and the CDW state
   is in the universality of the Ising phase transition.
The phase diagram for nonzero $J_\perp$ is shown 
   in Fig.\ \ref{fig:strong2}.
The CDW phase is realized when the condition (\ref{eq:CDW-Smott}) is
   satisfied.
We note that, within the strong-coupling expansion to second order,
   the CDW phase does not exist for $t_\parallel=t_\perp$.

Finally we discuss effects of the remaining interactions,
$H_{V_\parallel}$, $H_{V'}$, and $H_{\mathrm{pair}}$.
We find that we may ignore $H_{V_\parallel}$ and $H_{V'}$ since they
 yield only a constant
energy shift in the second-order perturbation theory.
By contrast, the pair-hopping term changes the phase boundary.
Since $H_{\mathrm{pair}}|+\rangle\!\rangle_j=0$ and
   $H_{\mathrm{pair}}|-\rangle\!\rangle_j=
    t_{\mathrm{pair}}|-\rangle\!\rangle_j$, 
   the interaction part of the Hamiltonian Eq.\ (\ref{eq:Hint_G})
   is modified as 
   $H_{\mathrm{int}}'=H_{\mathrm{int}}+H_{\mathrm{pair}}$, where
\begin{eqnarray}
H_{\mathrm{int}}'
&=& 
\left(
\begin{array}{cc}
V_\perp-\frac{3}{4}J_\perp   &  0 \\
0  &    U  + t_{\mathrm{pair}}
\end{array}
\right)
.
\end{eqnarray}
The main effect of $t_{\mathrm{pair}}$ is to change the coupling
   constant $h^z$ in Eq.\ (\ref{eq:Heff_Gaussian}) to
   $h^z=U-V_\perp+3J_\perp/4+t_{\mathrm{pair}}$.
In this case, the critical behavior is still governed by the Gaussian
   theory, and the critical point appears at
\begin{eqnarray}
U-V_\perp+ \frac{3}{4} J_\perp + t_{\mathrm{pair}} = 0
.
\label{eq:crit-DS2}
\end{eqnarray}
Thus, for $t_\mathrm{pair}>0$, the pair hopping term tends to
stabilize the D-Mott phase.
As shown in the last subsection, it also stabilizes the CDW phase,
   and the net effect of the
   pair hopping is to suppress the S-Mott phase sandwiched
   by the D-Mott and the CDW phases.

\subsection{SF state as AF ordering of rung-current and 
   SF--D-Mott transition}

\begin{figure}[t]
\includegraphics[width=8.cm]{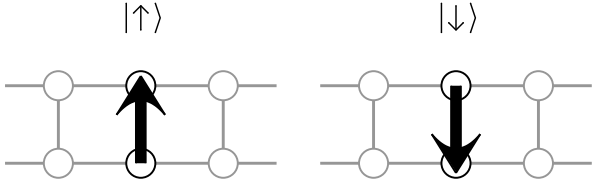}
\caption{
Schematic illustration of the states $|\!\uparrow\,\rangle$
   and $|\!\downarrow\,\rangle$.
The arrow denotes a state with a finite current running in
   the arrow's direction.
}
\label{fig:current}
\end{figure}

In this subsection, we study the SF state in the ladder system
   using the strong-coupling expansion.
Our starting point is the pair-hopping Hamiltonian 
   $H_{\mathrm{pair}}$ (\ref{eq:Hpair}).
The eigenstates of $H_\mathrm{pair}$ are given by $|1\rangle_j$,
  $|2\rangle_j$, $(|3\rangle_j+|4\rangle_j)/\sqrt{2}$, 
   and $(|3\rangle_j-|4\rangle_j)/\sqrt{2}$, satisfying
\begin{eqnarray}
H_{\mathrm{pair}} |1 \rangle_j
&=& 
H_{\mathrm{pair}} |2 \rangle_j =0
, \\
H_{\mathrm{pair}} \frac{| 3 \rangle_j - | 4 \rangle_j}{\sqrt{2}}
&=&
-t_\mathrm{pair} \, \frac{| 3 \rangle_j - | 4 \rangle_j}{\sqrt{2}}
, \\
H_{\mathrm{pair}} \frac{| 3 \rangle_j + | 4 \rangle_j}{\sqrt{2}}
&=&
+t_\mathrm{pair} \, \frac{| 3 \rangle_j + | 4 \rangle_j}{\sqrt{2}}
.
\end{eqnarray}
We thus find that the pair hopping term favors the on-site singlet
   state $(|3\rangle_j-|4\rangle_j)/\sqrt2$.
Anticipating competition between the on-site singlet state and the
   rung singlet state $(|1\rangle_j-|2\rangle_j)/\sqrt2$ that has an
   energy gain of $-3J_\perp/4$ from the exchange term $H_{J_\perp}$, 
   we will consider in this subsection the situation where
   $t_\mathrm{pair}\simeq3J_\perp/4$ and $J_\perp$ is the largest
   energy scale in the problem.
Introducing
   $\delta t_\mathrm{pair}=t_\mathrm{pair}-3J_\perp/4$
   $(|\delta t_\mathrm{pair}|\ll J_\perp)$,
   we define $\widetilde{H}_0$ and $\widetilde{H}'$ by
\begin{eqnarray}
\widetilde{H}_0
&=&
H_{J_{\perp}} + H_{\mathrm{pair}}^{(0)}
,
\\
\widetilde{H}'
&=&
H_{U} + H_{V_\perp} + H_{t_\parallel} + H_{t_\perp}
 + H_{\mathrm{pair}}'
,
\end{eqnarray}
   where 
   $H_{\mathrm{pair}}^{(0)}$ and $H_\mathrm{pair}'$ are obtained from
   $H_\mathrm{pair}$ by replacing $t_\mathrm{pair}$ with $3J_\perp/4$
   and $\delta t_\mathrm{pair}$, respectively.
The unperturbed Hamiltonian $\widetilde{H}_0$ has eigenstates,
\begin{eqnarray}
\widetilde{H}_0 \,  \frac{| 1 \rangle_j - | 2 \rangle_j}{\sqrt{2}}
&=& 
-\frac{3}{4}J_\perp \, \frac{| 1 \rangle_j - | 2 \rangle_j}{\sqrt{2}}
, \\
\widetilde{H}_0 \, \frac{| 3 \rangle_j - | 4 \rangle_j}{\sqrt{2}}
&=&
-\frac{3}{4}J_\perp \, \frac{| 3 \rangle_j - | 4 \rangle_j}{\sqrt{2}}
, \\
\widetilde{H}_0 \, \frac{| 1 \rangle_j + | 2 \rangle_j}{\sqrt{2}}
&=& 
+ \frac{1}{4}J_\perp \, 
  \frac{| 1 \rangle_j + | 2 \rangle_j}{\sqrt{2}}
, \\
\widetilde{H}_0 \, \frac{| 3 \rangle_j + | 4 \rangle_j}{\sqrt{2}}
&=&
+ \frac{3}{4}J_\perp \, \frac{| 3 \rangle_j + | 4 \rangle_j}{\sqrt{2}}
.
\end{eqnarray}
We will focus on the degenerate low-energy states
   $(| 1 \rangle_j - | 2 \rangle_j)/\sqrt2$
    and $(| 3 \rangle_j - | 4 \rangle_j)/\sqrt2$
   and work with the following
   states that break time reversal symmetry,
\begin{eqnarray}
|\!\uparrow\, \rangle_j &\equiv& 
\frac{1}{2} \bigl[\bigl(|1\rangle_j - |2 \rangle_j\bigr)
               +i \bigl(|3\rangle_j - |4 \rangle_j\bigr)
            \bigr]
,
\\
|\!\downarrow\, \rangle_j &\equiv& 
\frac{1}{2} \bigl[\bigl(|1\rangle_j - |2 \rangle_j\bigr)
               -i \bigl(|3\rangle_j - |4 \rangle_j\bigr)
            \bigr]
.
\end{eqnarray}
We regard them as states with finite current running on
   the $j$th rung (Fig.\ \ref{fig:current}), as
   they are eigenstates of the ``rung-current operator'' defined by
\begin{eqnarray}
\hat{J}_j
&\equiv& 
i\sum_\sigma
  \left( 
    c_{j,1,\sigma}^\dagger \, c^{}_{j,2,\sigma} 
 -  c_{j,2,\sigma}^\dagger \, c^{}_{j,1,\sigma} 
  \right)
\end{eqnarray}
   with eigenvalues $\pm2$,
\begin{eqnarray}
\hat{J}_j |\!\uparrow\,\rangle_j =+2\, |\!\uparrow\,\rangle_j
,\hspace*{.5cm}
\hat{J}_j |\!\downarrow\,\rangle_j =-2\, |\!\downarrow\,\rangle_j
.
\end{eqnarray}
We note that $\hat{J}$ is not a true current operator for
$\widetilde{H}_0$ due to the pair hopping term.

\begin{figure}[t]
\includegraphics[width=7.cm]{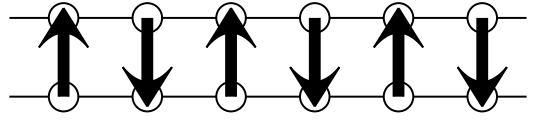}
\caption{
Staggered flux state described as a N\'eel
   ordered state of the pseudo-spin
   states, $|\!\uparrow\,\rangle$ and $|\!\downarrow\,\rangle$.
}
\label{fig:SF}
\end{figure}

The SF state has a long-range alternating order of 
   $|\!\uparrow\,\rangle$ and $|\!\downarrow\,\rangle$ or,
   equivalently, of currents circulating around each plaquette
  (Fig.\ \ref{fig:SF}).\cite{Fjaerestad}
To verify the existence of the SF phase, we derive a low-energy
   effective theory, in perturbation expansion in $H'$, for
   the low-energy states
    $|\!\uparrow\,\rangle_j$ and $|\!\downarrow\,\rangle_j$, 
   which we regard as up and down states of a pseudo-spin. 
In this picture, the antiferromagnetic ordering of the
   pseudo-spins corresponds to the staggered flux phase.
The lowest-order contribution in $\widetilde{H}'$ comes from the
   nonvanishing matrix elements in the subspace of
   $|\!\uparrow\,\rangle_j$ and $|\!\downarrow\,\rangle_j$,
\begin{eqnarray}
\langle\,\uparrow\!|\widetilde{H}'|\!\uparrow\,\rangle_j
=
\langle\,\downarrow\!|\widetilde{H}'|\!\downarrow\,\rangle_j
\!\!&=&\!\!
\frac{1}{2}(U+V_\perp - \delta t_\mathrm{pair}),
\\
\langle\,\uparrow\!|\widetilde{H}'|\!\downarrow\,\rangle_j
=
\langle\,\downarrow\!|\widetilde{H}'|\!\uparrow\,\rangle_j
\!\!&=&\!\!
-\frac{1}{2}(U-V_\perp- \delta t_\mathrm{pair}),
\qquad
\end{eqnarray}
from which we obtain the first-order effective Hamiltonian
\begin{equation}
H^{(1)}_\mathrm{SF}
=
 - \frac{1}{2}\left(U-V_\perp- \delta t_{\mathrm{pair}}\right)
    \sum_j \tilde\sigma^x_j
 +  \mathrm{const.}
,
\label{eq:H^1_SF}
\end{equation}
where $\tilde\sigma^a_j$ are the Pauli matrices ($a=x,y,z$).
The lowest-order contributions in $t_\parallel$ and $t_\perp$ come
   from the second-order processes,
\begin{eqnarray}
H^{(2a)}_\mathrm{SF} &=&
 H_{t_\parallel} \, \frac{1}{\widetilde{E}_0-\widetilde{H}_{0}} \,
             H_{t_\parallel}
,\\
H^{(2b)}_\mathrm{SF} &=&
 H_{t_\perp} \, \frac{1}{\widetilde{E}_0-\widetilde{H}_{0}} \,
             H_{t_\perp},
\end{eqnarray}
   where $\widetilde{E}_0=-3J_\perp N/4$ with $N$ being the number of
   rungs in the system. 
The nonzero matrix elements of $H^{(2a)}_\mathrm{SF}$ are given by
\begin{equation}
\langle\, \uparrow,\downarrow\!| H^{(2a)}_\mathrm{SF}
   |\! \uparrow,\downarrow\,\rangle_j
=
\langle\, \downarrow,\uparrow\!| H^{(2a)}_\mathrm{SF}
   |\! \downarrow,\uparrow\,\rangle_j
=
-\frac{8t_{\parallel}^2}{3J_\perp}
,
\end{equation}
where $|\mu,\nu\rangle_j\equiv|\mu\rangle_j|\nu\rangle_{j+1}$
($\mu,\nu=\uparrow,\downarrow$).
We can thus write $H^{(2a)}_\mathrm{SF}$ as
\begin{equation}
H^{(2a)}_\mathrm{SF}
=
\frac{4t_{\parallel}^2}{3J_\perp}
\sum_{j} \left(\tilde\sigma_j^z \, \tilde\sigma_{j+1}^z -1 \right)
.
\label{eq:H^2a_SF}
\end{equation}
On the other hand, the nonzero matrix elements of
$H^{(2b)}_\mathrm{SF}$ are
\begin{eqnarray}
&&
\langle\,\uparrow\!|H^{(2b)}_\mathrm{SF}|\!\uparrow\,\rangle_j
=
\langle\,\downarrow\!|H^{(2b)}_\mathrm{SF}|\!\downarrow\,\rangle_j
\nonumber\\
&&
=
\langle\,\uparrow\!|H^{(2b)}_\mathrm{SF}|\!\downarrow\,\rangle_j
=
\langle\,\downarrow\!|H^{(2b)}_\mathrm{SF}|\!\uparrow\,\rangle_j
=
-\frac{4t_\perp^2}{3J_\perp},
\qquad
\end{eqnarray}
from which we obtain
\begin{equation}
H^{(2b)}_\mathrm{SF}=
-\frac{4t_\perp^2}{3J_\perp}
\sum_j\tilde\sigma^x_j +\mathrm{const}.
\label{eq:H^2b_SF}
\end{equation}
From Eqs.\ (\ref{eq:H^1_SF}), (\ref{eq:H^2a_SF}), and
(\ref{eq:H^2b_SF}), we find that the total effective Hamiltonian is
the Ising chain in a transverse field,
\begin{equation}
H^\mathrm{eff}_\mathrm{SF}
=
\sum_{j} 
\left(
\widetilde{K} \, \tilde\sigma_j^z \, \tilde\sigma_{j+1}^z
 -\tilde h  \, \tilde\sigma_j^x
\right)
,
\end{equation}
   where the antiferromagnetic exchange coupling $\widetilde K$ and
   the magnitude of the transverse field $\tilde h$ are given by
\begin{equation}
\widetilde{K}
=
\frac{4t_\parallel^2}{3J_\perp}
,\,\,\,\,
\tilde h = \frac{1}{2}\left(U-V_\perp- \delta t_{\mathrm{pair}}
                     +\frac{8t_\perp^2}{3J_\perp}\right)
.
\label{eq:SF_K-h}
\end{equation}
This model exhibits an Ising criticality at $\widetilde K=|\tilde h|$:
   the N\'eel ordered phase ($\widetilde K>|\tilde h|$) corresponds to
   the SF phase, 
   while for $\widetilde K < |\tilde h|$ the system is disordered.
The disordered ground state for $\tilde h>\widetilde K>0$ is continuously
   connected with the ground state at $\tilde h\to\infty$, i.e.,
   the eigenstate of $\tilde\sigma^x$ with eigenvalue $+1$.
This state corresponds to the D-Mott state in the original Hubbard
   ladder, since
\begin{eqnarray}
|{\tilde\sigma^x\!=+1}\rangle_j &\!\!=\!\!&
\frac{1}{\sqrt{2}}
\left(|\!\uparrow\,\rangle_j + |\!\downarrow\,\rangle_j \right)
\nonumber \\
&\!\!=\!\!&
\frac{1}{\sqrt{2}}
\left(|1\rangle_j - |2\rangle_j \right)
\to |\mbox{D-Mott} \rangle
.
\label{eq:sigmax}
\end{eqnarray}
Hence we conclude that the Ising disordered phase corresponds to the
   D-Mott phase.

It is interesting to rewrite the transverse magnetic field
$\tilde h$ as
\begin{eqnarray}
\tilde h &=&
 \frac{1}{2} \left(U-V_\perp + \frac{3}{4} J_\perp 
   -t_{\mathrm{pair}}
   +\frac{8t_\perp^2}{3J_\perp}\right).
\end{eqnarray}
The SF phase is realized when the inequality
\begin{equation}
-\frac{16t^2}{3J_\perp}<
U-V_\perp+\frac{3}{4}J_\perp-t_\mathrm{pair}<0
\end{equation}
is satisfied
(assuming $t_\parallel=t_\perp=t$), where
we have to keep in mind the assumption that
$t_\mathrm{pair}\approx\frac{3}{4}J_\perp$.

\section{Weak-Coupling Approach}\label{sec:weakcoupling}

In this section, we study the phase diagram of the generalized Hubbard
ladder, treating the two-particle interactions as weak perturbations.
To diagonalize the single-particle hopping Hamiltonian, we define the
Fourier transform,
   $c_{j,\sigma}({k_\perp\!=\!0})
     =(c_{j,1,\sigma}+c_{j,2,\sigma})/\sqrt{2}$,
   $c_{j,\sigma}({k_\perp\!=\!\pi})
     =(c_{j,1,\sigma}-c_{j,2,\sigma})/\sqrt{2}$, and
   $c_\sigma(\bm{k})=\sum_j e^{-ikj}
                      c_{j,\sigma}(k_\perp)/\sqrt{N}$,
where $\bm{k}=(k,k_\perp)$ and the lattice spacing $a$ is set equal
to 1.
The kinetic energy term then becomes
\begin{equation}
H_0
\equiv
H_{t_\parallel}+H_{t_\perp}=
\sum_{\bm{k},\sigma}
\varepsilon(\bm{k}) \, 
  c_{\sigma}^\dagger (\bm{k}) \,   c_{\sigma} (\bm{k}) ,
\end{equation}
   where
   $\varepsilon(\bm{k}) = -2 t_\parallel \cos k - t_\perp \cos k_\perp$.
For $t_\perp< 2t_\parallel$, both the bonding ($k_\perp=0$) and
   antibonding ($k_\perp=\pi$) energy bands are partially filled,
   and their Fermi points are located at $k=\pm k_{F,k_\perp}$ with
   $k_{F,0}=\frac{\pi}{2}+\delta$ and $k_{F,\pi}=\frac{\pi}{2}-\delta$,
   where $\delta\equiv \sin^{-1} (t_\perp/2t_\parallel)$.
At these Fermi points the Fermi velocity takes the common value
   $v_{F} = 2t_\parallel [1-( t_\perp/2t_\parallel )^2 ]^{1/2}$.
In the following analysis
   we restrict ourselves to the isotropic hopping case 
   $t_\parallel=t_\perp(\equiv t)$.

\subsection{Order parameters}

Let us first define order parameters characterizing insulating phases
studied in this section.
We consider the CDW, SF, $p$-density-wave (PDW), and
   $f$-density-wave (FDW) states as possible density-wave ordered
   states.
Their order parameters are written as
\begin{eqnarray}
O_{\mathrm{A}} \!\!&=&\!\! \frac{1}{2N}
\sum_{\bm{k},\sigma} f_\mathrm{A}(\bm{k}) \, 
   c_{\sigma}^\dagger(\bm{k}) \, c_{\sigma}(\bm{k}+\bm{Q})
\nonumber\\
&\equiv&\!\!
\frac{1}{N}\sum_j(-1)^j\mathcal{O}_\mathrm{A}(j),
\end{eqnarray}
   where $\bm{Q}=(\pi,\pi)$ and A = CDW,
   SF, PDW, FDW.
The form factor $f_\mathrm{A}(\bm{k})$ are given by 
  $f_\mathrm{CDW} = 1$, $f_\mathrm{SF} = \cos k- \cos k_\perp$,
  $f_\mathrm{PDW} = \sin k$, and
  $f_\mathrm{FDW} = \sin k \, \cos k_\perp$.
Order parameters for the spin density waves are 
  not considered, since their correlations decay exponentially in the
   bulk of the phase diagram of our model.
It is clear that the CDW order parameter,
\begin{equation}
\mathcal{O}_\mathrm{CDW}=\frac{1}{2}(n_{j,1}-n_{j,2}),
\end{equation}
has nonvanishing average in the CDW states (\ref{eq:CDW1}) and
(\ref{eq:CDW2}).
The order parameter of the SF state is
\begin{equation}
\mathcal{O}_\mathrm{SF}=
\frac{1}{4i} \hat{J}_{\mathrm{P},j},
\end{equation}
   where the operator $\hat{J}_{\mathrm{P},j}$ denotes a current 
   circulating around a plaquette:
\begin{eqnarray}
\hat{J}_{\mathrm{P},j}
\!\!&\equiv & \!\!
i \sum_\sigma
\Bigl(
  c_{j,1,\sigma}^\dagger \, c^{}_{j,2,\sigma}
+ c_{j,2,\sigma}^\dagger \, c^{}_{j+1,2,\sigma} 
\nonumber \\ && {}
+ c_{j+1,2,\sigma}^\dagger \, c^{}_{j+1,1,\sigma}
+ c_{j+1,1,\sigma}^\dagger \, c^{}_{j,1,\sigma} 
-\mathrm{H.c.}
\Bigr).
\qquad
\end{eqnarray}
The PDW phase is a Peierls dimerized state along the
   leg direction with inter-leg phase difference $\pi$,
 characterized by the order parameter,
\begin{equation}
\mathcal{O}_\mathrm{PDW}=
 \frac{i}{4}
\sum_\sigma
\biggl(
c_{j+1,1,\sigma}^\dagger \, c_{j,1,\sigma}^{}
-c_{j+1,2,\sigma}^\dagger \, c_{j,2,\sigma}^{}
+\mathrm{H.c.}
\biggr).
\end{equation}
The FDW state is a different kind of staggered current states.
Its order parameter is
\begin{equation}
\mathcal{O}_\mathrm{FDW}=\frac{1}{4}
 \left(\hat{J}_{+,j}-\hat{J}_{-,j}\right),
\end{equation}
where the operators $\hat{J}_{\pm,j}$ represent currents flowing
along the diagonal directions of plaquettes:
\begin{eqnarray}
\hat{J}_{+,j}\!\!&=&\!\!
i\sum_\sigma
\left(c^\dagger_{j+1,2,\sigma}c^{}_{j,1,\sigma}
     -c^\dagger_{j,1,\sigma}c^{}_{j+1,2,\sigma}\right),\\
\hat{J}_{-,j}\!\!&=&\!\!
i\sum_\sigma
\left(c^\dagger_{j+1,1,\sigma}c^{}_{j,2,\sigma}
     -c^\dagger_{j,2,\sigma}c^{}_{j+1,1,\sigma}\right).
\end{eqnarray}
The long-range order of staggered currents flowing 
   along diagonals of the plaquettes has been examined in a spinless
   ladder system.\cite{Nersesyan1991}

We also introduce order parameters of the
   $s$-wave and $d$-wave superconductivity,
\begin{eqnarray}
O_\mathrm{A}
= \frac{1}{2N}
\sum_{\bm{k}} f_\mathrm{A}(\bm{k}) \, 
      c_{\uparrow}(\bm{k}) \, c_{\downarrow}(-\bm{k}) ,
\end{eqnarray}
   where A = SCs and SCd, and
   $f_\mathrm{SCs}=1$ and
   $f_\mathrm{SCd}=\cos k-\cos k_\perp$.

\subsection{Bosonization}

We bosonize the Hubbard ladder Hamiltonian in this subsection.
Following the standard bosonization scheme, we linearize the energy
bands around the Fermi points.
The linearized kinetic energy is given by
\begin{equation}
H_0
=
\sum_{\bm{k},p,\sigma}
v_F (pk-k_{F,k_\perp})  \,
  c_{p,\sigma}^\dagger (\bm{k}) \, c^{}_{p,\sigma} (\bm{k})
,
\end{equation}
   where the index $p=+/-$ denotes the right/left-moving electron.
We introduce field operators of the right- and left-going
electrons defined by
\begin{subequations}
\begin{eqnarray}
\psi_{p,\sigma,+}(x)\!\!&=&\!\!
\frac{1}{\sqrt{L}}\sum_k e^{ikx}c_{p,\sigma}(k,0), \\
\psi_{p,\sigma,-}(x)\!\!&=&\!\!
\frac{1}{\sqrt{L}}\sum_k e^{ikx}c_{p,\sigma}(k,\pi),
\end{eqnarray}
\end{subequations}
where $L$ is the length of the system: $L=Na$.
The linearized kinetic energy now reads
\begin{equation}
H_0=
v_F\int dx \sum_{p,\sigma,\zeta}
\psi^\dagger_{p,\sigma,\zeta}
\left(-ip\frac{d}{dx}-k_{F,k_\perp}\right)
\psi^{}_{p,\sigma,\zeta},
\end{equation}
where $k_\perp=0$ ($\pi$) for $\zeta=+$ ($-$).

The interactions among low-energy excitations near the Fermi points,
 $H_I=H_\mathrm{int}+H_{V_\parallel}+H_{V'}+H_\mathrm{pair}$,
are written as $H_I=\int dx\mathcal{H}_I$, where
\begin{eqnarray}
\mathcal{H}_I
&\!\!=\!\!&
   \frac{1}{4}
   \sum_{p,\sigma}{\sum_{\zeta_i=\pm}}'
\biggl[
   g_{1\parallel}^{\epsilon\bar\epsilon} \,
      \psi_{p,\sigma,\zeta_1}^\dagger \,
      \psi_{-p,\sigma,\zeta_2}^\dagger \,
      \psi_{p,\sigma,\zeta_4} \,
      \psi_{-p,\sigma,\zeta_3}
\nonumber \\&& \hspace*{13mm} {}
  + g_{1\perp}^{\epsilon\bar\epsilon} \,
      \psi_{p,\sigma,\zeta_1}^\dagger \,
      \psi_{-p,-\sigma,\zeta_2}^\dagger \,
      \psi_{p,-\sigma,\zeta_4} \,
      \psi_{-p,\sigma,\zeta_3}
\nonumber \\&& \hspace*{13mm} {}
  + g_{2\parallel}^{\epsilon\bar\epsilon} \,
      \psi_{p,\sigma,\zeta_1}^\dagger \,
      \psi_{-p,\sigma,\zeta_2}^\dagger \,
      \psi_{-p,\sigma,\zeta_4} \,
      \psi_{p,\sigma,\zeta_3}
\nonumber \\&& \hspace*{13mm} {}
  + g_{2\perp}^{\epsilon\bar\epsilon} \,
      \psi_{p,\sigma,\zeta_1}^\dagger \,
      \psi_{-p,-\sigma,\zeta_2}^\dagger \,
      \psi_{-p,-\sigma,\zeta_4} \,
      \psi_{p,\sigma,\zeta_3}
\nonumber \\&& \hspace*{13mm} {}
  + g_{3\parallel}^{\epsilon\bar\epsilon} \,
      \psi_{p,\sigma,\zeta_1}^\dagger \,
      \psi_{p,\sigma,\zeta_2}^\dagger \,
      \psi_{-p,\sigma,\zeta_4} \,
      \psi_{-p,\sigma,\zeta_3}
\nonumber \\&& \hspace*{13mm} {}
  + g_{3\perp}^{\epsilon\bar\epsilon} \,
      \psi_{p,\sigma,\zeta_1}^\dagger \,
      \psi_{p,-\sigma,\zeta_2}^\dagger \,
      \psi_{-p,-\sigma,\zeta_4} \,
      \psi_{-p,\sigma,\zeta_3}
\biggr]
.
\nonumber \\
\label{eq:Hint_g-ology}
\end{eqnarray} 
Here $\epsilon=\zeta_1\zeta_3$ and $\bar\epsilon=\zeta_1\zeta_2$.
The primed summation over $\zeta_i$ ($i=1,\ldots,4$) is taken
   under the condition 
   $\zeta_1 \zeta_2 \zeta_3 \zeta_4 = +1$,
   which comes from the momentum conservation condition 
   in the transverse direction. 
The coupling constants $g_{i\parallel}^{\epsilon\bar\epsilon}$
 and $g_{i\perp}^{\epsilon\bar\epsilon}$
 are related to the
original coupling constants in the Hamiltonian (\ref{eq:H}):
\begin{eqnarray}
\frac{g_{i\parallel}^{\epsilon\bar\epsilon}}{a}
\!\!&=&\!\!
  l_\epsilon V_\perp
+ \frac{l_\epsilon}{4} J_\perp
+ m_{i,\epsilon} V_\parallel
+ l_\epsilon m_{i,\epsilon} V'
,
\label{eq:gpara}
\\
\frac{g_{i\perp}^{\epsilon\bar\epsilon}}{a}
\!\!&=&\!\!
U
+ l_\epsilon  V_\perp
+ \frac{l_{\epsilon,\bar\epsilon}}{4} J_\perp
+ l_{\bar\epsilon}  t_{\mathrm{pair}}
+ m_{i,\epsilon} V_\parallel
+ l_\epsilon m_{i,\epsilon} V'
\nonumber\\
\label{eq:gperp}
\end{eqnarray} 
   with the numerical factors defined by
   $l_\pm = \pm 1$, 
   $l_{\pm,+}=\mp 3$, $l_{\pm,-}=\pm 1$.
   $m_{1,+}=m_{3,+}=-1$, $m_{1,-}=m_{3,-}=-2$,
   $m_{2,+}=+2$, $m_{2,-}=+1$.
We have neglected the so-called $g_4$ terms describing
   the forward scattering processes within the same branch
   (left-/right-mover), since including these terms would only cause
   nonuniversal quantitative differences to the ground state phase
   diagram.
In Eqs.\ (\ref{eq:gpara}) and (\ref{eq:gperp}),
   we have estimated the coupling constants 
   in lowest order in the interaction of the Hubbard model. 
The higher-order contributions can play a crucial role of changing
   topology of a phase diagram, if different kinds of quantum
   criticalities accidentally occur simultaneously when lowest-order
   coupling constants are used, as is the case in the 1D
   extended Hubbard model at half-filling.\cite{Tsuchiizu2002}
This is not the case in the ladder model of our interest, and 
   we will use the lowest order form, Eqs.\ (\ref{eq:gpara}) and
   (\ref{eq:gperp}). 

We apply the Abelian bosonization method\cite{Emery,Solyom,Gogolin_book}
and rewrite the kinetic energy in terms of bosonic fields:
$H_0=\int dx\mathcal{H}_0$, where
\begin{equation}
{\cal H}_0
=
\frac{v_F}{2\pi} 
 \sum_{\nu=\rho,\sigma}\sum_{r=\pm}
\left[
  \left(\pi \Pi_{\nu r}\right)^2 
  + \left(\frac{d \phi_{\nu r}}{dx}\right)^2
\right]
.
\end{equation}
Here the suffices $\rho$ and $\sigma$ refer to the charge and spin 
sectors and $r=\pm$ refer to the even and odd sectors.
The operator $\Pi_{\nu r}(x)$ is a canonically conjugate variable 
   to $\phi_{\nu r}(x)$ and satisfies
$[\phi_{\nu r}(x),\Pi_{\nu' r'}(x')] = i \, \delta(x-x') \, 
  \delta_{\nu,\nu'} \, \delta_{r,r'}$.
We then introduce chiral bosonic fields
\begin{eqnarray}
\phi_{\nu r}^{\pm}(x)
\equiv
\frac{1}{2}
\left[
  \phi_{\nu r}(x) \mp \pi \int_{-\infty}^{x} dx' \, \Pi_{\nu r}(x')
\right]
,
\end{eqnarray} 
   which satisfy the commutation relations
   $[\phi_{\nu r}^{\pm}(x),\phi_{\nu' r'}^{\pm}(x')] =
    \pm i(\pi/4) \, \mathrm{sgn}(x-x') \, 
    \delta_{\nu,\nu'} \, \delta_{r,r'}$
   and
   $[\phi_{\nu r}^+(x),\phi_{\nu' r'}^-(x')] =
     i(\pi/4) \, 
     \delta_{\nu,\nu'} \, \delta_{r,r'}$.
The right-moving and left-moving chiral fields $\phi^+(x,\tau)$
   and $\phi^-(x,\tau)$ are functions of $\tau-i(x/v_F)$ and
   $\tau+i(x/v_F)$, respectively, where $\tau$ is imaginary time.
The kinetic-energy density can also be written as
\begin{equation}
{\cal H}_0
= \frac{v_F}{\pi} \sum_{p=\pm}\sum_{\nu=\rho,\sigma}\sum_{r=\pm}
        \left( \frac{d\phi_{\nu r}^p}{dx} \right)^2
.
\label{eq:H_harmonic}
\end{equation}
We also introduce the field $\theta_{\nu r}$ defined by
  $\theta_{\nu r}=\phi_{\nu r}^+ - \phi_{\nu r}^-$.
The $\theta$ field satisfies the commutation relation
   $[ \phi_{\nu r}(x), \theta_{\nu' r'}(x') ] =
   -i \pi \Theta(-x+x')\delta_{r,r'}$,
   where $\Theta(x)$ is the Heaviside step function.

To express the electron fields in terms of the bosons, we define a new
set of chiral bosonic fields
\begin{equation}
\varphi_{p,s,\zeta}
  =
     \phi_{\rho +}^p
   + \zeta  \phi_{\rho -}^p
   + s \phi_{\sigma +}^p
   + s \zeta \phi_{\sigma -}^p
,
\end{equation}
    where $p=\pm$, $s=\pm$, and $\zeta=\pm$.
The chiral bosons obey the commutation relations
   $[\varphi_{p,s,\zeta}(x),\varphi_{p,s',\zeta'}(x')]$
    $= ip\pi \, \mathrm{sgn}(x-x') \, 
    \delta_{s,s'}\,\delta_{\zeta,\zeta'}$
   and
   $[\varphi_{+,s,\zeta},\varphi_{-,s',\zeta'}]
    = i\pi \,\delta_{s,s'}\,\delta_{\zeta,\zeta'}$.

The field operators of the right- and left-moving electrons
are then written as
\begin{equation}
\psi_{p,\sigma,\zeta} =
 \frac{\eta_{\sigma,\zeta}}{\sqrt{2\pi a} }
\exp \left( ipk_{F,k_\perp} x 
 + i p\, \varphi _{p,s,\zeta} \right),
\label{eq:field} 
\end{equation}
   where $s=+$ for $\sigma=\uparrow$ and $s=-$ for
   $\sigma=\downarrow$. 
The Klein factors $\eta_{\sigma,\zeta}$, which satisfy
   $\{\eta_{\sigma,\zeta},\eta_{\sigma',\zeta'}\}
   =2\delta_{\sigma,\sigma'}\delta_{\zeta,\zeta'}$,
   are introduced in order to retain the correct anticommutation
   relation of the field operators between different spin and
   the band index.
From Eq.\ (\ref{eq:field}) the density operator is given by
\begin{equation}
\rho_{p,\sigma,\zeta}(x)
   = \, 
   :\! \psi_{p,\sigma,\zeta}^\dagger\, \psi_{p,\sigma,\zeta}^{} \! :
   \, = \,
   \frac{1}{2\pi} \, \frac{d}{dx} \varphi_{p,s,\zeta}(x)
.
\label{eq:density}
\end{equation}
The Hamiltonian and the order parameters contain 
   only products of the Klein factors such
   as\cite{Schulz1996,Fjaerestad}
   $\Gamma \equiv
   \eta_{\uparrow,+} \, \eta_{\downarrow,+} \, 
       \eta_{\uparrow,-} \, \eta_{\downarrow,-} $,
   $h_\sigma \equiv \eta_{\sigma,+} \, \eta_{\sigma,-}$,
   and 
   $h'_\zeta \equiv \eta_{\uparrow,\zeta} \, \eta_{\downarrow,\zeta}$,
   which satisfy
   $ \Gamma = -h_\uparrow \, h_\downarrow = + h'_+ \, h'_- $.
Since $\Gamma^2=+1$, $h^2=(h')^2=-1$,
   the eigenvalues are $\Gamma=\pm 1$,
   $h=\pm i$, and $h'=\pm i$.
We will adopt the following convention:
 $\Gamma= + 1$, $h_\sigma = i$, $h_\zeta'= i\zeta$.

In the bosonized Hamiltonian the phase field $\phi_{\rho-}$ appears
   in the form $\cos (2\phi_{\rho-}+4\delta x)$ with
   $\delta=\sin^{-1}(t_\perp/2t_\parallel)$.
Since $t_\perp$ $(=t_\parallel)$ is not small,
   we can safely assume that the $\delta$ is relevant 
   and the electrons are not confined in the legs.
   \cite{Tsuchiizu1999,LeHur,Tsuchiizu2001}
In this case the $\cos (2\phi_{\rho-}+4\delta x)$ terms 
    become irrelevant.
We thus discard them as well as
    other terms with higher-order scaling dimensions.
The interaction term Eq.\ (\ref{eq:Hint_g-ology}) reduces to
\begin{eqnarray}
\mathcal{H}_I
&\!\!=\!\!& 
\sum_{\nu = \rho,\sigma} \sum_{r=\pm}
\frac{g_{\nu r}}{2\pi^2} 
   \left(\partial_x \phi_{\nu r}^+ \right)
   \left(\partial_x \phi_{\nu r}^- \right)
\nonumber \\
&&\!\!{} + \frac{1}{2\pi^2 a^2}\Bigl[ 
    g_{c+, \overline{c-}} \,
    \cos 2 \phi_{\rho+} \,
    \cos 2 \theta_{\rho-} 
\nonumber \\ && {}
+   g_{c+, s+} \,
    \cos 2 \phi_{\rho+} \,
    \cos 2 \phi_{\sigma+}   
  \nonumber \\
&& {}
+   g_{c+, s-} \, 
    \cos 2 \phi_{\rho+} \,
    \cos 2 \phi_{\sigma-}   
  \nonumber \\
&& {}
+   g_{c+, \overline{s-}} \,
    \cos 2 \phi_{\rho+} \,\,
    \cos 2 \theta_{\sigma-}  
  \nonumber \\
&& {}
+   g_{\overline{c-},s+} \,
    \cos 2 \theta_{\rho-} \,
    \cos 2 \phi_{\sigma+}
  \nonumber \\
&& {}
+   g_{\overline{c-},s-}\,
    \cos 2 \theta_{\rho-} \,
    \cos 2 \phi_{\sigma-} 
  \nonumber \\
&& {}
+   g_{\overline{c-},\overline{s-}}\,
    \cos 2 \theta_{\rho-} \,\,
    \cos 2 \theta_{\sigma-}
\nonumber \\
&& {}
+   g_{s+, s-}\,
    \cos 2 \phi_{\sigma+} \,
    \cos 2 \phi_{\sigma-}   
  \nonumber \\
&& {}
+   g_{s+,\overline{s-}}\,
    \cos 2 \phi_{\sigma+}\,
    \cos 2 \theta_{\sigma-}   
\Bigr]
,
\label{eq:Hint}
\end{eqnarray}
   where the coupling constants 
   for the bilinear terms of the density operators are given by
\begin{subequations}
\begin{eqnarray}
g_{\rho+} \!\!&=&\!\! \sum_{\epsilon=\pm}
   ( g_{2\parallel}^{+\epsilon}+g_{2\perp}^{+\epsilon}
    -g_{1\parallel}^{\epsilon\epsilon}), \\
g_{\rho-} \!\!&=&\!\! \sum_{\epsilon=\pm} \epsilon
   ( g_{2\parallel}^{+\epsilon}+g_{2\perp}^{+\epsilon}
    -g_{1\parallel}^{\epsilon\epsilon}), \\
g_{\sigma+} \!\!&=&\!\! \sum_{\epsilon=\pm}
   ( g_{2\parallel}^{+\epsilon}-g_{2\perp}^{+\epsilon}
    -g_{1\parallel}^{\epsilon\epsilon}), \\
g_{\sigma-} \!\!&=&\!\! \sum_{\epsilon=\pm} \epsilon
   ( g_{2\parallel}^{+\epsilon}-g_{2\perp}^{+\epsilon}
    -g_{1\parallel}^{\epsilon\epsilon}),
\end{eqnarray}
\end{subequations}
   and the coupling constants for
   the nonlinear terms are given by
\begin{subequations}
\begin{eqnarray}
g_{c+,\overline{c-}} \!\!&=&\!\!  - g_{3\perp}^{-+}, \\
g_{c+,s+}\!\!&=&\!\! - g_{3\parallel}^{+-} + g_{3\parallel}^{--}, \\
g_{c+,s-}\!\!&=&\!\! - g_{3\perp}^{+-}, \\
g_{c+, \overline{s-}}\!\!&=&\!\! + g_{3\perp}^{--}, \\
g_{\overline{c-},s+}\!\!&=&\!\! - g_{1\perp}^{-+}, \\
g_{\overline{c-},s-}\!\!&=&\!\! - g_{2\perp}^{-+}, \\
g_{\overline{c-},\overline{s-}}\!\!&=&\!\!
    + g_{2\parallel}^{-+} - g_{1\parallel}^{-+}, \\
g_{s+,s-}\!\!&=&\!\! + g_{1\perp}^{++}, \\
g_{s+,\overline{s-}}\!\!&=&\!\! + g_{1\perp}^{--}.
\end{eqnarray}
\end{subequations}
We note that the umklapp scattering (the $g_3$ terms) generates cosine
potentials that lock the $\phi_{\rho+}$ field.

The coupling constants in Eq.\ (\ref{eq:Hint}) are not independent
   parameters.
Imposing the global spin-rotation SU(2) symmetry on the interaction
   terms Eq.\ (\ref{eq:Hint_g-ology}), we find that the relations
\begin{subequations}
\begin{eqnarray}
&& g_{2\parallel}^{++} - g_{2\perp}^{++} - g_{1\parallel}^{++}
 + g_{1\perp}^{++} =0, \\
&& g_{2\parallel}^{+-} - g_{2\perp}^{+-} - g_{1\parallel}^{--}
 + g_{1\perp}^{--} =0, \\
&& g_{2\parallel}^{--} - g_{2\perp}^{--} - g_{1\parallel}^{+-}
 + g_{1\perp}^{+-} =0,
\label{eq:G_SU(2)}
\\
&& g_{2\parallel}^{-+} - g_{2\perp}^{-+} - g_{1\parallel}^{-+}
 + g_{1\perp}^{-+} =0, \\
&& g_{3\parallel}^{+-} - g_{3\parallel}^{--}
      - g_{3\perp}^{+-} + g_{3\perp}^{--} =0,
\end{eqnarray}
\end{subequations}
must hold.
In terms of the coupling constants in Eq.\ (\ref{eq:Hint}),
   these relations read
\begin{subequations}
\begin{eqnarray}
g_{\sigma+}+g_{\sigma-} + 2g_{s+,s-} = 0,&&
\label{eq:su2_sigma}
\\
g_{\sigma+}-g_{\sigma-} + 2g_{s+,\overline{s-}} = 0,&&
\\
g_{\overline{c-},s+} - g_{\overline{c-},s-}
-g_{\overline{c-},\overline{s-}}
 = 0,&&
\label{eq:su2_c-}
\\
 g_{c+,s+} - g_{c+,s-} - g_{c+,\overline{s-}} = 0.&&
\label{eq:su2_c+}
\end{eqnarray}
\label{eq:su2's}%
\end{subequations}
We have ignored Eq.\ (\ref{eq:G_SU(2)}) which is the constraint on the 
   irrelevant cosine term $\propto\cos(2\phi_{\rho-}+4\delta x)$.
Since the SU(2) symmetry of the original Hubbard Hamiltonian
   (\ref{eq:H}) cannot be broken, the coupling constants in 
   Eq.\ (\ref{eq:Hint}) must satisfy
   Eqs.\ (\ref{eq:su2_sigma})-(\ref{eq:su2_c+}) in the course of
   renormalization.

Finally, the order parameters are written in terms of the phase fields:
\begin{subequations}
\begin{eqnarray}
&&
\mathcal{O}_{\mathrm{CDW}}
\propto
  \cos \phi_{\rho+} \,
  \sin \theta_{\rho-} \,
  \cos \phi_{\sigma+} \,
  \cos \theta_{\sigma-} 
\nonumber \\ && \qquad\qquad
{} -
  \sin \phi_{\rho+} \,
  \cos \theta_{\rho-} \,
  \sin \phi_{\sigma+} \,
  \sin \theta_{\sigma-} ,
\label{eq:order_CDW}
\\
&&
\mathcal{O}_{\mathrm{SF}}
\propto
  \cos \phi_{\rho+} \,
  \cos \theta_{\rho-} \,
  \cos \phi_{\sigma+} \,
  \cos \theta_{\sigma-} 
\nonumber \\ && \qquad\quad
{} +
  \sin \phi_{\rho+} \,
  \sin \theta_{\rho-} \,
  \sin \phi_{\sigma+} \,
  \sin \theta_{\sigma-} ,
\label{eq:order_SF}
\\
&&
\mathcal{O}_{\mathrm{PDW}}
\propto
  \cos \phi_{\rho+} \,
  \cos \theta_{\rho-} \,
  \sin \phi_{\sigma+} \,
  \sin \theta_{\sigma-} 
\nonumber \\ && \qquad\qquad
{} +
  \sin \phi_{\rho+} \,
  \sin \theta_{\rho-} \,
  \cos \phi_{\sigma+} \,
  \cos \theta_{\sigma-} ,
\\
&&
\mathcal{O}_{\mathrm{FDW}}
\propto
  \cos \phi_{\rho+} \,
  \sin \theta_{\rho-} \,
  \sin \phi_{\sigma+} \,
  \sin \theta_{\sigma-} 
\nonumber \\ && \qquad\qquad
{}  -
  \sin \phi_{\rho+} \,
  \cos \theta_{\rho-} \,
  \cos \phi_{\sigma+} \,
  \cos \theta_{\sigma-} .\qquad\quad
\\
&&
\mathcal{O}_\mathrm{SCd}
\propto
  e^{i \theta_{\rho+}}
  \cos \theta_{\rho-} \,
  \cos \phi_{\sigma+} \,
  \cos \phi_{\sigma-}
\nonumber \\ && \qquad\qquad
{}  - i \,
  e^{i \theta_{\rho+}}
  \sin \theta_{\rho-} \,
  \sin \phi_{\sigma+} \,
  \sin \phi_{\sigma-} 
,
\\ 
&&
\mathcal{O}_\mathrm{SCs}
\propto
  e^{i \theta_{\rho+}}
  \cos \theta_{\rho-} \,
  \sin \phi_{\sigma+} \,
  \sin \phi_{\sigma-}
\nonumber \\ && \qquad\qquad
{}  - i \,
  e^{i \theta_{\rho+}}
  \sin \theta_{\rho-} \,
  \cos \phi_{\sigma+} \,
  \cos \phi_{\sigma-} 
.
\label{eq:O_SCs}
\end{eqnarray}
\label{order-parameters}
\end{subequations}

\subsection{Critical properties in the charge and spin modes}

In this subsection, we study the ground state phase diagram through 
   qualitative analysis of the bosonized Hamiltonian (\ref{eq:Hint}).
First we classify the phases that can appear at half-filling, and
   then discuss (a) the Gaussian criticality in the charge sector and
   (b) the Ising and SU(2)$_2$ criticalities in the spin sector.

\subsubsection{Classification of phases}

\begin{table*}
\caption{Pattern of phase locking.
The $*$ symbol indicates that a bosonic field is not locked.
$I_i$s are integers.
}
\label{table:phase-locking}
\begin{ruledtabular}
\begin{tabular}{lccccc}
Phase &
 $\langle\phi_{\rho+}\rangle$ & $\langle\theta_{\rho-}\rangle$ &
 $\langle\phi_{\sigma+}\rangle$ & $\langle\phi_{\sigma-}\rangle$ &
 $\langle\theta_{\sigma-}\rangle$ \\
\hline
CDW &
 $\frac{\pi}{2}I_0+\pi I_1$ & $\frac{\pi}{2}(I_0+1)+\pi I_2$ &
 $\frac{\pi}{2}I_0+\pi I_3$ & $*$ & $\frac{\pi}{2}I_0+\pi I_4$ \\
SF &
 $\frac{\pi}{2}I_0+\pi I_1$ & $\frac{\pi}{2}I_0+\pi I_2$ &
 $\frac{\pi}{2}I_0+\pi I_3$ & $*$ & $\frac{\pi}{2}I_0+\pi I_4$ \\
PDW &
 $\frac{\pi}{2}(I_0+1)+\pi I_1$ & $\frac{\pi}{2}(I_0+1)+\pi I_2$ &
 $\frac{\pi}{2}I_0+\pi I_3$ & $*$ & $\frac{\pi}{2}I_0+\pi I_4$ \\
FDW &
 $\frac{\pi}{2}(I_0+1)+\pi I_1$ & $\frac{\pi}{2}I_0+\pi I_2$ &
 $\frac{\pi}{2}I_0+\pi I_3$ & $*$ & $\frac{\pi}{2}I_0+\pi I_4$ \\
S-Mott &
 $\frac{\pi}{2}I_0+\pi I_1$ & $\frac{\pi}{2}(I_0+1)+\pi I_2$ &
 $\frac{\pi}{2}I_0+\pi I_3$ & $\frac{\pi}{2}I_0+\pi I_4$ & $*$ \\
D-Mott &
 $\frac{\pi}{2}I_0+\pi I_1$ & $\frac{\pi}{2}I_0+\pi I_2$ &
 $\frac{\pi}{2}I_0+\pi I_3$ & $\frac{\pi}{2}I_0+\pi I_4$ & $*$ \\
S'-Mott &
 $\frac{\pi}{2}(I_0+1)+\pi I_1$ & $\frac{\pi}{2}(I_0+1)+\pi I_2$ &
 $\frac{\pi}{2}I_0+\pi I_3$ & $\frac{\pi}{2}I_0+\pi I_4$ & $*$ \\
D'-Mott &
 $\frac{\pi}{2}(I_0+1)+\pi I_1$ & $\frac{\pi}{2}I_0+\pi I_2$ &
 $\frac{\pi}{2}I_0+\pi I_3$ & $\frac{\pi}{2}I_0+\pi I_4$ & $*$ \\
\end{tabular}
\end{ruledtabular}
\end{table*}

In general all the modes become massive in the extended Hubbard ladder
at half-filling.
This means that in the bosonized Hamiltonian (\ref{eq:Hint})
cosine terms are relevant at low energies and that the bosonic phase
fields are locked at some fixed values (integer multiples of $\pi/2$)
where the relevant cosine potentials are minimized.\cite{Lin}
The locked phase fields can be treated as classical variables, and the 
average value of an order parameter is found by 
substituting the locked phases into Eq.~(\ref{order-parameters}).
A nonvanishing order parameter signals which phase is realized.
We can reverse the logic and find the configuration of the locked phase
fields for each insulating phase by imposing its order parameter to
have its maximum modulus.
This is what we do in the following analysis.

In the SF, CDW, PDW, and FDW phases the ground state breaks a Z$_2$
symmetry.
Therefore the order parameter of these phases can have a
nonvanishing value at zero temperature even in one dimension.
In each phase the bosonic fields $\phi_{\rho+}$, $\theta_{\rho-}$,
$\phi_{\sigma+}$, and $\theta_{\sigma-}$ are pinned at a point where
the modulus of the corresponding order parameter is maximized.
From Eq.~(\ref{order-parameters}) we can easily find at which values
the bosonic fields are locked for the four phases.
The result is summarized in Table~\ref{table:phase-locking}.

Once the configuration of locked phase fields is understood for the SF
and the CDW phases, we can also find that for the 
D-Mott and the S-Mott phases using the following arguments.
On the one hand, we know from the strong-coupling analysis that these
two insulating phases are Ising disordered phases of the SF and the
CDW phases, respectively, where
the $\theta_{\sigma-}$ field is locked.
On the other hand, the Hamiltonian (\ref{eq:Hint}) has some cosine
potentials that can lock the $\phi_{\sigma-}$ field.
Since the $\phi_{\sigma-}$ field is a conjugate field to
$\theta_{\sigma-}$, these two fields cannot be locked at the same
time.
In fact, it is known \cite{Schulz1996} that an Ising
phase transition must be associated with switching of phase locking
from one bosonic field to its conjugate field.
We can thus obtain the D-Mott and the S-Mott phases from the SF and
the CDW phases by exchanging the role of the $\phi_{\sigma-}$ field
and the $\theta_{\sigma-}$ field, 
arriving at the phase locking pattern shown in Table
\ref{table:phase-locking}.
A brief comment on the connection to the superconducting states is
in order here.
If we ignore the $\rho+$ mode for the moment, the order parameter of
the $d$-wave ($s$-wave) superconductivity takes nonzero amplitude
when the locked phases ($\langle\theta_{\rho-}\rangle$,
$\langle\phi_{\sigma+}\rangle$,
and $\langle\phi_{\sigma-}\rangle$)
of the D-Mott (S-Mott) phase are substituted into
$\mathcal{O}_\mathrm{SCd(s)}$.
This is consistent with the previous
results\cite{Fabrizio,Noack,Sigrist,Tsunetsugu1994,Khveshchenko1994,%
         Nagaosa,Dagotto,Schulz1996,Balents1996,%
         Orignac,Tsuchiizu2001}
that, upon doping, the D-Mott state turns into the $d$-wave
superconducting state in the $t$-$J$ or Hubbard ladder.
The effect of carrier doping is to make the umklapp term irrelevant
and to leave the $\phi_{\rho+}$ field unlocked.
The operator $e^{i\theta_{\rho+}}$ representing the superconducting
correlation then becomes quasi-long-range ordered.

It is possible to construct a disorder parameter that characterizes
the Ising transitions and that has a nonvanishing expectation value in 
the D-Mott and the S-Mott phases.
A candidate operator for the disorder parameter is
\begin{eqnarray}
\mu_j\!\!&=&\!\!
\exp\left(
i\frac{\pi}{2}\sum_{i=1}^j
X_i
\right),\nonumber\\
X_i\!\!&=&\!\!
c_{i,1,\uparrow}^\dagger c_{i,2,\uparrow}^{}
      +c_{i,2,\uparrow}^\dagger c_{i,1,\uparrow}^{}
\nonumber\\
&&
     {} -c_{i,1,\downarrow}^\dagger c_{i,2,\downarrow}^{}
      -c_{i,2,\downarrow}^\dagger c_{i,1,\downarrow}^{}
.
\label{eq:mu_j}
\end{eqnarray}
In the weak-coupling limit we take the continuum limit and express
the operator (\ref{eq:mu_j}) in terms of the bosonic fields.
We then obtain
\begin{equation}
\mu_j=\exp[i\phi_{\sigma-}(j)].
\label{eq:mu_j_boson}
\end{equation}
Indeed, the disorder parameter $\mu_j$ takes a nonzero value in the
D-Mott and the S-Mott phases where the $\phi_{\sigma-}$ field is locked.
In the strong-coupling limit studied in Sec.\ III, we may impose the
condition that
$n_{i,1}+n_{i,2}=2$ and $S^z_{i,1}+S^z_{i,2}=0$ on every rung.
Under this condition we find that
$\exp(i\frac{\pi}{2}X_i)=1-\frac{1}{2}X_i^2$
and $\mu_j$ reduces to
\begin{eqnarray}
\mu_j\!\!&=&\!\!
\prod^j_{i=1}\biggl[
\left(
c^\dagger_{i,1,\uparrow}c^\dagger_{i,1,\downarrow}
c_{i,2,\downarrow}^{}c_{i,2,\uparrow}^{}+\mathrm{H.c.}\right)
\nonumber\\
&&\quad
-\left(S^+_{i,1}S^-_{i,2}+S^-_{i,1}S^+_{i,2}\right)
\biggr],
\end{eqnarray}
which acts on the pseudo-spin states defined in Secs.\ IIIA and IIIC as
$\mu_j|+\rangle_i=|-\rangle_i$ and
$\mu_j|\!\!\!\uparrow\,\rangle_i
 =|\!\!\!\downarrow\,\rangle_i$ for $i\le j$.
This means that we can write $\mu_j=\prod^j_i\tau^x_i$ and
$\mu_j=\prod^j_i\tilde\sigma^x_i$ near the CDW--S-Mott and the SF--D-Mott
transitions, respectively.
They are indeed the disorder parameter of the quantum Ising
model\cite{Gogolin_book} that
describes the CDW--S-Mott and the SF--D-Mott Ising transitions.

Since the PDW and the FDW phases break Z$_2$ symmetry, we can
naturally expect that these two phases should also have their own
Ising disordered phases.
We shall call them S'-Mott and D'-Mott phases for the reason that will 
become clear below.
The configuration of phase locking in the S'-Mott and D'-Mott phases
can be obtained from that of the PDW and FDW phases by exchanging
$\langle\phi_{\sigma-}\rangle$ and $\langle\theta_{\sigma-}\rangle$;
see Table \ref{table:phase-locking}.
We see immediately that the phase-locking pattern of the S'-Mott
(D'-Mott) state differs from that of the S-Mott (D-Mott) only
in the locking of the $\phi_{\rho+}$ field shifted by $\pi/2$.
This implies that the phase transition between S'-Mott (D'-Mott) state 
and the S-Mott (D-Mott) state is a Gaussian transition in the
$\phi_{\rho+}$ mode, and that the S'-Mott (D'-Mott) state should
evolve into the $s$-wave ($d$-wave) superconducting state upon carrier 
doping as in the S-Mott (D-Mott) state.

The nature of the S'-Mott state can be deduced through
its similarity to the S-Mott state (\ref{eq:S-Mott}).
We first note that, as mentioned above, the S'-Mott state is related
to the S-Mott state by a $\pi/2$ shift of the $\phi_{\rho+}$ mode,
which is equivalent to translation by half unit cell, in such a way
that the PDW state is related to the CDW state.
This suggests that the center of mass of a singlet in the S'-Mott state 
should be located at a center of a plaquette.
Noting that $\cos k\cos k_\perp$ is positive ($s$-wave like) at all
the Fermi points,
$\bm{k}=\biglb(\pm(\frac{\pi}{2}+\delta),0\bigrb)$ and
$\biglb(\pm(\frac{\pi}{2}-\delta),\pi\bigrb)$,
of the ladder model, we speculate that the singlet-pair wave function
(or the symmetry of a Cooper pair in the $s$-wave superconducting
state realized upon doping) is of the form
$\cos k\cos k_\perp
 c_\uparrow^\dagger(\bm{k})c_\downarrow^\dagger(-\bm{k})$
in momentum space.
In real space this corresponds to a linear combination of two singlets
formed between diagonal sites of a plaquette.
From these consideration we come to propose the following wave
function as a representative of the S'-Mott state:
\begin{eqnarray}
|\mbox{S'-Mott}\rangle\!\!&=&\!\!
\prod_j\frac{1}{2}(
c_{j,1,\uparrow}^\dagger c_{j+1,2,\downarrow}^\dagger
-c_{j,1,\downarrow}^\dagger c_{j+1,2,\uparrow}^\dagger
\nonumber\\
&&\quad {}
+c_{j,2,\uparrow}^\dagger c_{j+1,1,\downarrow}^\dagger
-c_{j,2,\downarrow}^\dagger c_{j+1,1,\uparrow}^\dagger
)|0\rangle.
\qquad
\label{eq:S'-Mott}
\end{eqnarray}
This state mostly consists of singlets along the diagonal direction
of plaquettes but also contains resonating singlets that are formed by
two spins on different legs that can be separated far away.

The D'-Mott state consists of singlets that would turn into $d$-wave
Cooper pairs upon doping.
Since the singlet-pair wave function in the D-Mott state is
$\cos k_\perp$ in momentum space, we expect that the singlet pairs in
the D'-Mott state should be of the form $\cos k$.
In real space this corresponds to a linear combination of singlets
formed in the leg direction.
This leads to the following wave function
\begin{equation}
|\mbox{D'-Mott}\rangle=\prod_j\left[
\sum_{l=1,2}
\frac{
c_{j,l,\uparrow}^\dagger c_{j+1,l,\downarrow}^\dagger
-c_{j,l,\downarrow}^\dagger c_{j+1,l,\uparrow}^\dagger
}{2}
\right]|0\rangle
\label{eq:D'-Mott}
\end{equation}
as a representative of the D'-Mott state.
It is easy to see by expanding the product that this state is a resonating
valence bond state in which some singlets can be formed out of two spins
that are separated arbitrary far away along a leg.
However, amplitude of the states having such a long-distance singlet is
exponentially suppressed with the distance between the two spins.

It is interesting to note that the wave function (\ref{eq:S'-Mott})
can be constructed from the S-Mott wave function (\ref{eq:S-Mott}) by
replacing $c^\dagger_{j,l,\sigma}$ with
$c^\dagger_{j+1,\bar{l},\sigma}$, where $\bar{l}=2$ (1) for $l=1$ (2)
such that
$c^\dagger_{j,l,\uparrow}c^\dagger_{j,l,\downarrow}\to
 (c^\dagger_{j+1,\bar{l},\uparrow}c^\dagger_{j,l,\downarrow}
 +c^\dagger_{j,l,\uparrow}c^\dagger_{j+1,\bar{l},\downarrow})/\sqrt2$.
This rule can also be used to construct the wave function of the
D'-Mott state (\ref{eq:D'-Mott}) from that of the D-Mott state
(\ref{eq:D-Mott}).

Since the $\phi_{\sigma-}$ field is locked in the S'-Mott and D'-Mott
phases, the operator (\ref{eq:mu_j_boson}) also serves as the disorder
parameter in the PDW--S'-Mott and the FDW--D'-Mott transitions of the
Ising universality class.
In fact, the disorder parameter (\ref{eq:mu_j_boson}) takes a nonzero 
value in any of the Mott phases and vanishes otherwise.

\begin{figure}[t]
\includegraphics[width=8.5cm]{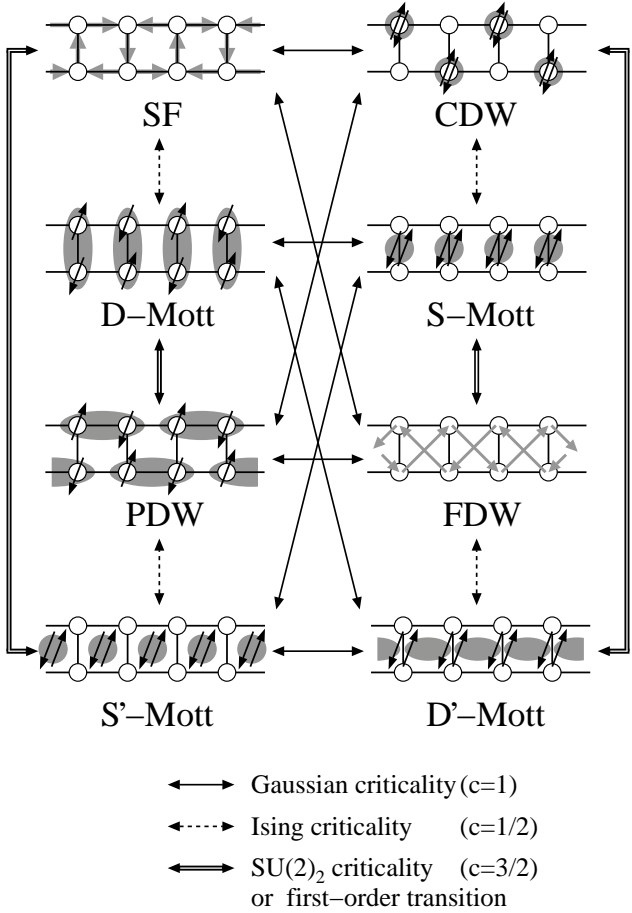}
\caption{
Schematic illustration of the phase diagram 
   under the global SU(2) symmetry.
The phase transitions indicated by the solid
   (dashed) arrows are the $c=1$ ($c=1/2$) criticality.
The phase transitions indicated by the double arrows are
  either the $c=3/2$ SU(2)$_2$ criticality or first order;
  see discussion in Sec.\ IVC3 and Fig.\ \ref{fig:phase-g}.
The diagonal solid arrows denote the Gaussian transitions
in the $\phi_{\rho+}$ mode.
}
\label{fig:diagram}
\end{figure}

The various insulating phases and phase transitions among them are
schematically shown in Fig.\ \ref{fig:diagram}.
In this figure phase transitions between a phase in the left column
and another in the right column, such as transitions between the Mott
phases, are the $c=1$ Gaussian criticality.
It would be interesting to find an order parameter that can distinguish
different Mott phases.
The transitions in the vertical direction within a column 
are, if continuous, either
the $c=1/2$ Ising criticality or the $c=3/2$ SU(2)$_2$ criticality.
The latter may be replaced by a first-order transition.
We will discuss these transitions in more detail in the following
subsubsections.  

A brief comment on the related earlier works is in order here.
The top four phases (SF, CDW, S-Mott, and D-Mott) in
Fig.~\ref{fig:diagram} and the Gaussian and 
Ising transitions between these phases
have been found in the weak-coupling RG analysis of the SO(5)
symmetric ladder model by Lin, Balents, and
Fisher.\cite{Lin}
The misidentification of the SF phase with the PDW phase made in
this work has been corrected later by Fj{\ae}restad and
Marston.\cite{Fjaerestad}
We have pointed out the existence of four more phases in the
generalized Hubbard ladder model
and determined the universality
class of the phase transitions between all the 8 phases.

\subsubsection{Gaussian criticality in the charge degrees of freedom}

First we discuss the Gaussian criticality when all the modes except
the relative charge mode ($\rho-$) become massive at some higher energy
scale.
This situation is relevant for the horizontal transitions in
Fig.~\ref{fig:diagram}: SF--CDW, D-Mott--S-Mott, PDW--FDW, and 
S'-Mott--D'-Mott transitions.
We take the D-Mott--S-Mott phase transition as an example.
Without loss of generality we may assume that
   the phase variables are locked at 
   $\langle\phi_{\rho+}\rangle=\langle\phi_{\sigma+}\rangle
    =\langle\phi_{\sigma-}\rangle=0$ mod $\pi$.
Below the energy scale at which the three fields are locked,
   we can replace the cosine terms in the Hamiltonian 
   Eq.\ (\ref{eq:Hint}) by their average:
   $\cos 2\phi_{\rho+} \to 
    c_{\rho+} \equiv \langle \cos 2\phi_{\rho+}\rangle$,
   $\cos 2\phi_{\sigma+} \to 
    c_{\sigma+} \equiv \langle \cos 2\phi_{\sigma+}\rangle$, and
   $\cos 2\phi_{\sigma-}\to 
    c_{\sigma-} \equiv \langle \cos 2\phi_{\sigma-}\rangle$,
   where $c_{\rho+}$, $c_{\sigma+}$, and $c_{\sigma-}$ are
   nonuniversal positive constants that depend on bare interactions.
We then have the effective theory 
\begin{eqnarray}
\mathcal{H}_{\rho-}
\!\!&=&\!\!
\frac{v_F}{\pi}
\left[
(\partial_x\phi^+_{\rho-})^2 + (\partial_x\phi^-_{\rho-})^2
\right]
\nonumber\\
&&\!
{} +
\frac{g_{\rho-}}{2\pi^2}
  \left( \partial_x \phi_{\rho-}^+ \right)
  \left( \partial_x \phi_{\rho-}^- \right)
\nonumber\\
&&\!
{} +
\frac{g_{\overline{c-}}}{2\pi^2 a^2} \, \cos 2\theta_{\rho-}
,
\label{eq:Heff_gaussian}
\end{eqnarray}
   where the coupling constant $g_{\overline{c-}}$ is given by
\begin{equation}
g_{\overline{c-}}
=
c_{\rho+} \, g_{c+,\overline{c-}} +
c_{\sigma+} \, g_{\overline{c-},s+} +
c_{\sigma-} \, g_{\overline{c-},s-}
.
\end{equation}
Since the canonical dimension of $\cos 2\theta_{\rho-}$
   is $1$, the $g_{\overline{c-}}$ term is a relevant 
   perturbation and hence the system always becomes massive
   except when $g_{\overline{c-}}=0$.
If $g_{\overline{c-}}>0$, then the phase field is locked as
   $\langle \theta_{\rho-}\rangle=\pi/2$ mod $\pi$, which corresponds
   to the S-Mott phase.
When $g_{\overline{c-}}<0$, the phase field is locked as
   $\langle \theta_{\rho-}\rangle=0$ mod $\pi$, and 
   the ground state in this case turns out to be the D-Mott state.
The Gaussian criticality with the central charge $c=1$ is realized at 
   $g_{\overline{c-}}=0$.
In terms of the original Hubbard interactions
    the coupling constant $g_{\overline{c-}}$
    is given by
\begin{equation}
\frac{g_{\overline{c-}}}{a}
=
-C
\left(
U-V_\perp + \frac{3}{4}J_\perp + t_\mathrm{pair}
\right)
+C'(V_\parallel-V')
,
\label{eq:gc-}
\end{equation}
   where $C \equiv c_{\rho+}+c_{\sigma+}+c_{\sigma-}$ and
   $C'\equiv 2c_{\rho+}+2c_{\sigma+}-c_{\sigma-}$ are
   nonuniversal \emph{positive} constants.
Thus, the D-Mott (S-Mott) state appears
   when $U-V_\perp + 3J_\perp/4 + t_\mathrm{pair}
          -C'(V_\parallel-V')/C >0$ ($<0$),
   and the Gaussian criticality shows up at
\begin{eqnarray}
U-V_\perp + \frac{3}{4}J_\perp + t_\mathrm{pair}
 - \frac{C'}{C}(V_\parallel-V') = 0,
\label{eq:critical_G}
\end{eqnarray}
   which is the same as the phase boundary obtained from
   the strong-coupling analysis, Eq.\ (\ref{eq:crit-DS2}),
   for $V_\parallel=V'=0$.

The SF--CDW phase transition can be analyzed in a similar way.
We consider a situation where
   the phase variable $\theta_{\sigma-}$,
   instead of $\phi_{\sigma-}$, is locked at 
   $\langle\theta_{\sigma-}\rangle=0$ mod $\pi$.
In this case we can replace the cosine factor in the Hamiltonian as 
   $\cos2\theta_{\sigma-}\to c_{\overline{\sigma-}} \equiv
    \langle \cos 2\theta_{\sigma-}\rangle >0$.
The effective theory is given by Eq.\ (\ref{eq:Heff_gaussian}) with
   the coupling constant
$g_{\overline{c-}} =
   c_{\rho+} \, g_{c+,\overline{c-}} +
   c_{\sigma+} \, g_{\overline{c-},s+} +
   c_{\overline{\sigma-}} \, g_{\overline{c-},\overline{s-}}$.
The SF (CDW) state is realized for $g_{\overline{c-}}<0$ 
   $(>0)$, where the phase $\theta_{\rho-}$ is locked at 
   $0$ ($\pi/2$) mod $\pi$.
In terms of the original Hubbard interactions,
    the coupling constant $g_{\overline{c-}}$
    is given by Eq.\ (\ref{eq:gc-}) with
   $C=c_{\rho+}+c_{\sigma+}>0$ and
   $C'=2c_{\rho+}+2c_{\sigma+}+3c_{\overline{\sigma-}}$.
We thus conclude that the SF (CDW) state appears 
   for $U-V_\perp + \frac{3}{4}J_\perp + t_\mathrm{pair}
         - C'(V_\parallel-V')/C
         >0$ ($<0$),
   and the condition for the Gaussian criticality is given by
   Eq.\ (\ref{eq:critical_G}).

The other transitions of the $c=1$ Gaussian criticality can also be
   analyzed in the same manner.
We note that in addition to the Gaussian criticality in the $\rho-$
   mode discussed above, there is another Gaussian criticality in the
   $\rho+$ mode that govern the SF--FDW, CDW--PDW,
   D-Mott--D'-Mott, and S-Mott--S'-Mott transitions.

\subsubsection{Z$_2$ $\times$ O(3) symmetry 
   in the spin degrees of freedom
   and the Ising and SU(2)$_2$ criticality}

Here we focus on the case where the masses of the two charge modes
   ($\rho\pm$) are larger than those of the spin modes ($\sigma\pm$).
Below the mass scale of the charge modes we may regard that the
   $\phi_{\rho+}$ and $\theta_{\rho-}$ fields are locked by cosine
   potentials.
The effective low-energy theory is obtained from Eq.\ (\ref{eq:Hint})
   by replacing $\cos2\phi_{\rho_+}$ and $\cos2\theta_{\rho-}$ by their
   average values $c_{\rho+}\equiv\langle\cos2\phi_{\rho+}\rangle$
   and $c_{\overline{\rho-}}\equiv\langle\cos2\theta_{\rho-}\rangle$:
\begin{eqnarray}
{\cal H}_\sigma
\! &=& \!
\frac{v_F}{\pi}
\left[
   \left(\partial \phi_{\sigma+}^+ \right)^2
 + \left(\partial \phi_{\sigma+}^- \right)^2
 + \left(\partial \phi_{\sigma-}^+ \right)^2
 + \left(\partial \phi_{\sigma-}^- \right)^2
\right]
\nonumber \\ &&{}
+
\frac{g_{\sigma+}}{2\pi^2}
   \left(\partial \phi_{\sigma+}^+ \right)
   \left(\partial \phi_{\sigma+}^- \right)
+
    \frac{g_{s+}}{2\pi^2 a^2}\,
    \cos 2 \phi_{\sigma+}   
\nonumber \\ &&{}
+
\frac{g_{\sigma-}}{2\pi^2}
   \left(\partial \phi_{\sigma-}^+ \right)
   \left(\partial \phi_{\sigma-}^- \right)
\nonumber \\ &&{}
+
    \frac{g_{s-}}{2\pi^2 a^2}\,
    \cos 2 \phi_{\sigma-}   
+
    \frac{g_{\overline{s-}}}{2\pi^2 a^2}\,
    \cos 2 \theta_{\sigma-}  
  \nonumber \\
&&{}
+
    \frac{g_{s+, s-}}{2\pi^2 a^2}\,
    \cos 2 \phi_{\sigma+} \,
    \cos 2 \phi_{\sigma-}   
\nonumber \\ &&{}
+
    \frac{g_{s+,\overline{s-}}}{2\pi^2 a^2}\,
    \cos 2 \phi_{\sigma+}\,
    \cos 2 \theta_{\sigma-}   
,
\label{eq:Heff_spsm2}
\end{eqnarray}
   where the coupling constants $g_{s+}$, $g_{s-}$, and 
   $g_{\overline{s-}}$ are given by
\begin{subequations}
\begin{eqnarray}
g_{s+} 
&\equiv& 
  c_{\rho+} \, g_{c+,s+} 
+ c_{\overline{\rho-}} \, g_{\overline{c-},s+}
,
\label{eq:gs+}
\\
g_{s-} 
&\equiv&
  c_{\rho+} \, g_{c+,s-} 
+ c_{\overline{\rho-}} \, g_{\overline{c-},s-} 
,
\\
g_{\overline{s-}} 
&\equiv&
  c_{\rho+} \, g_{c+,\overline{s-}} 
+ c_{\overline{\rho-}} \, g_{\overline{c-},\overline{s-}} 
.
\label{eq:gs-}
\end{eqnarray}
\label{eq:gs's}%
\end{subequations}
The coupling constants in Eq.\ (\ref{eq:Heff_spsm2}) 
   are not completely free parameters,
   since the system has the spin-rotational SU(2) symmetry.
From Eqs.\ (\ref{eq:su2's}) and (\ref{eq:gs's}),
   the constraints on the coupling constants read
\begin{subequations}
\begin{eqnarray}
&& g_{s+} - g_{s-} - g_{\overline{s-}} =0,
\label{constraint1}
\\ &&
g_{s+,s-} = -\frac{1}{2} (g_{\sigma+}+g_{\sigma-}),
\\ &&
g_{s+,\overline{s-}} = - \frac{1}{2} (g_{\sigma+}-g_{\sigma-}).
\end{eqnarray}
\label{eq:SU2}%
\end{subequations}
To appreciate the SU(2) symmetry in the effective theory
(\ref{eq:Heff_spsm2}), we fermionize it by introducing
   spinless fermion fields $\psi_{p,r}$ ($p=\pm$ and $r=\pm$):
\begin{equation}
\psi_{\pm,r}(x) = \frac{\eta_r}{\sqrt{2\pi a}}
  \, \exp\left[ \pm i \,2\phi_{\sigma r}^{\pm}(x)\right],
\end{equation}
   where the index $r=+(-)$ refers to the total (relative) degrees
   of freedom of spin mode, and $\{\eta_r,\eta_{r'}\}=2\delta_{r,r'}$.
The density operators are given by
   $ :\! \psi_{p,\pm}^\dagger \, \psi_{p,\pm} \!:  \, = 
   \partial_x  \phi_{\sigma \pm}^p/\pi $.
We then introduce the Majorana fermions $\xi^n$ ($n=1\sim 4$) by
\begin{eqnarray}
\psi_{p,+}
=
\frac{1}{\sqrt{2}} \left(\xi_{p}^1+i\xi_{p}^2\right),
\hspace*{.2cm}
\psi_{p,-}
=
\frac{1}{\sqrt{2}} \left(\xi_{p}^4+i\xi_{p}^3\right)
.
\end{eqnarray}
These fields satisfy the anticommutation relations
  $\{\xi_p^n(x),\xi_{p'}^{n'}(x')\}  =
   \delta(x-x') \, \delta_{p,p'} \, \delta_{n,n'}$.
With the help of the SU(2) constraints (\ref{eq:SU2}),
we rewrite the effective Hamiltonian in terms of the Majorana
fermions:
\begin{eqnarray}
\mathcal{H}_\sigma
\!\!&=&\!\! {}
-i\frac{v_F}{2} 
\left(
  \bm{\xi}_+ \cdot \partial_x \bm{\xi}_+
- \bm{\xi}_- \cdot \partial_x \bm{\xi}_-
\right)
 - i m_t \,
  \bm{\xi}_+ \cdot \bm{\xi}_-
\nonumber \\ && {}
-i\frac{v_F}{2}
\left(
 \xi_+^4 \, \partial_x \xi_+^4
- \xi_-^4 \, \partial_x \xi_-^4
\right)
- i m_s \,
   \xi_+^4 \, \xi_-^4
\nonumber \\ && {}
+\frac{g_{\sigma+}}{4} \, 
\left(
 \bm{\xi}_+ \cdot \bm{\xi}_-
\right)^2
+ \frac{g_{\sigma-}}{2}
 \left( \bm{\xi}_+ \cdot \bm{\xi}_- \right)
 \, \xi_+^4 \, \xi_-^4
, \qquad
\label{eq:Heff_spin}
\end{eqnarray}
where we have introduced
$\mbox{\boldmath $\xi$}_p=(\xi_p^1,\xi_p^2,\xi_p^3)$
and
\begin{equation}
m_t\equiv -\frac{g_{s+}}{2\pi a},
\qquad
m_s \equiv -\frac{g_{s-}-g_{\overline{s-}}}{2\pi a}.
\label{eq:mtms}
\end{equation}
Thus the effective theory for the spin sector becomes
   O(3)$\times$Z$_2$ symmetric, i.e.,
the four Majorana fermions are grouped into a singlet $\xi^4$ with
   mass $m_s$ and a triplet $\bm{\xi}$ with mass $m_t$.
We note that the O(3)$\times$Z$_2$ symmetry also appears in the
low-energy effective theory of the isotropic Heisenberg
ladder.\cite{Shelton,Nersesyan1997}
It is known that, when $m_s,m_t\ne0$, the quartic marginal terms lead
   to mass renormalization, $m_s \to \widetilde{m}_s$ and
$m_t \to \widetilde{m}_t$, where\cite{Shelton,Gogolin_book}
\begin{eqnarray}
\widetilde{m}_t \!\!&=&\!\! 
  m_t 
  + \frac{g_{\sigma+}}{2\pi v_F} m_t \ln \frac{\Lambda}{|m_t|}
  + \frac{g_{\sigma-}}{4\pi v_F} m_s \ln \frac{\Lambda}{|m_s|},
\qquad
\label{eq:mt_tilde}
\\
\widetilde{m}_s \!\!&=&\!\!
  m_s
  + \frac{3g_{\sigma-}}{4\pi v_F} m_t \ln \frac{\Lambda}{|m_t|}.
\label{eq:ms_tilde}
\end{eqnarray}
Here $\Lambda$ is a high-energy cutoff.
The effective theory then reduces to
\begin{eqnarray}
\mathcal{H}_\sigma
&\!\!=\!\!& {}
-i\frac{v_F}{2} 
\left(
  \bm{\xi}_+ \cdot \partial_x \bm{\xi}_+
- \bm{\xi}_- \cdot \partial_x \bm{\xi}_-
\right)
 - i \widetilde{m}_t \,
  \bm{\xi}_+ \cdot \bm{\xi}_-
\nonumber \\ && {}
-i\frac{v_F}{2}
\left(
 \xi_+^4 \, \partial_x \xi_+^4
- \xi_-^4 \, \partial_x \xi_-^4
\right)
- i \widetilde{m}_s \,
   \xi_+^4 \, \xi_-^4
.
\label{eq:Majorana Hamiltonian}
\end{eqnarray}
It immediately follows from Eq.\ (\ref{eq:Majorana Hamiltonian}) that 
   the Ising criticality with $c=1/2$ 
   emerges as $\widetilde{m}_s \to 0$.
On the other hand,
   the critical properties for the O(3) invariant sector
   ($\widetilde{m}_t\to 0$)
   is known to be described by the SU(2)$_2$
   Wess-Zumino-Novikov-Witten model
   with the central charge $c=3/2$. 
   \cite{Gogolin_book,Tsvelik1990}

Let us examine the critical behavior in more detail using
the scaling equations for the coupling constants appearing in
the effective Hamiltonian (\ref{eq:Heff_spin}):
\begin{subequations}
\begin{eqnarray}
\frac{dG_t}{dl}\!\!&=&\!\!
G_t+G_tG_{\sigma+}+\frac{1}{2}G_sG_{\sigma-},
\label{eq:dG_t/dl}\\
\frac{dG_s}{dl}\!\!&=&\!\!
G_s+\frac{3}{2}G_tG_{\sigma-},
\label{eq:dG_s/dl}\\
\frac{dG_{\sigma+}}{dl}\!\!&=&\!\!
\frac{1}{2}G_{\sigma+}^2+\frac{1}{2}G_{\sigma-}^2+2G_t^2,
\label{eq:dG_sigma+/dl}\\
\frac{dG_{\sigma-}}{dl}\!\!&=&\!\!
G_{\sigma+}G_{\sigma-}+2G_tG_s,
\label{dG_sigma-/dl}%
\end{eqnarray}
\end{subequations}
where $dl=da/a$, $G_t=-g_{s+}/2\pi v_F$,
$G_s=-(g_{s-}-g_{\overline{s-}})/2\pi v_F$, and
$G_{\sigma\pm}=g_{\sigma\pm}/2\pi v_F$.
The coupling $G_s$ and $G_t$ are relevant,
 while $G_{\sigma\pm}$ are marginal.
Within the one-loop RG we find 4 stable fixed points,
$(G_t^*,G_s^*,G_{\sigma+}^*,G_{\sigma-}^*)
 =(\pm\infty,\pm\infty,\infty,\infty)$ and
  $(\pm\infty,\mp\infty,\infty,-\infty)$,
which correspond to the 8 phases listed in Fig.\ \ref{fig:diagram} and 
Table \ref{table:phase}.
The Ising criticality is governed by the unstable fixed point
$(G_t^*,G_s^*,G_{\sigma+}^*,G_{\sigma-}^*)=(\pm\infty,0,\infty,0)$,
where the Majorana fermion $\xi^4$ is massless.
The unstable fixed point
$(G_t^*,G_s^*,G_{\sigma+}^*,G_{\sigma-}^*)=(0,\pm\infty,0,0)$
corresponds to the SU(2)$_2$ criticality since the triplet $\bm{\xi}$
becomes massless.
Finally, we find another kind of unstable fixed points
$(G_t^*,G_s^*,G_{\sigma+}^*,G_{\sigma-}^*)=(0,\pm\infty,\infty,0)$,
where all the modes are massive.
To understand the nature of these unstable fixed points, let us
assume
$(g_{s+},g_{\overline{s-}}-g_{s-},g_{\sigma+},g_{\sigma-})
 =(0,2\lambda_1,2\lambda_2,0)$, 
 where $\lambda_{1,2}$ are constants
 ($\lambda_1\ne0$, $\lambda_2>0$).
This, together with the SU(2) constraint (\ref{eq:SU2}), leads to
$g_{\overline{s-}}=-g_{s-}=\lambda_1$ and
$g_{s+,\overline{s-}}=g_{s+,s-}=-\lambda_2<0$.
In this case the cosine terms in $\mathcal{H}_\sigma$
(\ref{eq:Heff_spsm2}) become
\begin{eqnarray}
&&{}
-\frac{\lambda_1}{2\pi^2a^2}(\cos2\phi_{\sigma-}-\cos2\theta_{\sigma-})
\nonumber\\
&&{}
-\frac{\lambda_2}{2\pi^2a^2}\cos2\phi_{\sigma+}
 (\cos2\phi_{\sigma-}+\cos2\theta_{\sigma-}).
\qquad
\label{eq:first-order}
\end{eqnarray}
Suppose that $\lambda_1>0$ and
 $\langle\phi_{\rho+}\rangle=\langle\theta_{\rho-}\rangle=0$.
We then find that the potential (\ref{eq:first-order}) has degenerate
minima at, e.g.,
$(\langle\phi_{\sigma+}\rangle,\langle\phi_{\sigma-}\rangle,
\langle\theta_{\sigma-}\rangle)
=(0,0,*)$ and $(\frac{\pi}{2},*,\frac{\pi}{2})$, where $*$ means that
the phase field is not locked.
Since these minima correspond to the D-Mott and PDW phases,
the unstable fixed point describes a first-order
transition between the D-Mott and PDW phases, respectively.
Hence we conclude that the unstable fixed points
$(G_t^*,G_s^*,G_{\sigma+}^*,G_{\sigma-}^*)=(0,\pm\infty,\infty,0)$
correspond to a first-order phase transition.
The phase transition at which the renormalized triplet mass $G_t^*$
vanishes can be either SU(2)$_2$ criticality or first-order
transition, depending on the sign of $G_{\sigma+}$ 
  \cite{Shankar}.
The condition for the SU(2)$_2$ criticality is
  $G_t=0$ and $G_{\sigma+}<0$ below the energy scale where $G_s$
becomes of order 1.
On the other hand, the first-order transition is realized if
  $G_t=0$ and $G_{\sigma+}>0$.

\begin{table}[t]
\caption{
Signs of the fixed-point coupling constants and the masses
 ($m_g$, $\widetilde m_s$, $\widetilde m_t$) in various phases.
}
\label{table:phase}
\begin{ruledtabular}
\begin{tabular}{lcccc}
 Phase   & $(g_{\overline{c-}}^*,g_{s+}^*,
             g_{s-}^*,g_{\overline{s-}}^*,g_{\sigma+}^*,g_{\sigma-}^*)$
         & $m_g$ & $\widetilde{m}_s$ & $\widetilde{m}_t$  \\ \hline 
 CDW     & $(+,-,0,-,+,-)$ &  $+$  &     $-$       &    $+$    \\
 SF      & $(-,-,0,-,+,-)$ &  $-$  &     $-$       &    $+$    \\
 PDW     & $(-,+,0,+,+,-)$ &  $-$  &     $+$       &    $-$    \\
 FDW     & $(+,+,0,+,+,-)$ &  $+$  &     $+$       &    $-$    \\
 S-Mott  & $(+,-,-,0,+,+)$ &  $+$  &     $+$       &    $+$    \\
 D-Mott  & $(-,-,-,0,+,+)$ &  $-$  &     $+$       &    $+$    \\
 S'-Mott  & $(-,+,+,0,+,+)$ &  $-$  &     $-$       &    $-$    \\
 D'-Mott  & $(+,+,+,0,+,+)$ &  $+$  &     $-$       &    $-$ 
\end{tabular}
\end{ruledtabular}
\end{table}

The phase fields are locked at some multiples of $\pi/2$
depending on signs of the relevant coupling constants at a fixed point,
$(g_{\overline{c-}}^*,g_{s+}^*,g_{s-}^*,g_{\overline{s-}}^*)$,
of the cosine potentials in 
Eqs.\ (\ref{eq:Heff_gaussian}) and (\ref{eq:Heff_spsm2}).
Comparing the configuration of the locked phases and those listed in
Table \ref{table:phase-locking}, we can find out to which phase
the ground state belongs for given combination of the renormalized
coupling constants,
$(g_{\overline{c-}}^*,g_{s+}^*,g_{s-}^*,g_{\overline{s-}}^*)$.
Table \ref{table:phase} summarizes for each phase the signs of
these renormalized coupling constants including 
$g_{\sigma\pm}^*$, which is positive (negative) when 
$\phi_{\sigma\pm}$ ($\theta_{\sigma\pm}$) is locked.
When writing Table \ref{table:phase}, we have used the fact (a) that
either one of $g_{s-}^*$ and $g_{\overline{s-}}^*$ must vanish
except at the Ising criticality because $\phi_{\sigma-}$
and $\theta_{\sigma-}$ are conjugate fields, and (b) that
Eq.\ (\ref{constraint1}) constraints possible combinations of signs
of $g_{s+}$, $g_{s-}$, and $g_{\overline{s-}}$.

The coupling constants listed in Table \ref{table:phase} also
   determine the signs of masses $m_g(=g_{\overline{c-}}/2\pi a)$, 
   $\widetilde m_s$, and $\widetilde m_t$ through
   Eqs.\ (\ref{eq:mtms}), (\ref{eq:mt_tilde}), and (\ref{eq:ms_tilde}).
The Gaussian ($c=1$), Ising ($c=1/2$), and SU(2)$_2$ ($c=3/2$) 
   criticalities are realized when
   $m_g=0$, $\widetilde{m}_s=0$, and $\widetilde{m}_t=0$, respectively. 
From Table \ref{table:phase} we can therefore figure out which
criticality can occur at each phase transition where the relevant
mass changes sign.
The universality class of the phase transitions is also summarized in
Fig.\ \ref{fig:diagram}.
We find from Table \ref{table:phase} that the CDW--S-Mott and
SF--D-Mott phase transitions
   are indeed in the Ising universality class and
   the D-Mott--S-Mott phase transition is in the Gaussian 
   universality class, in agreement with
   the strong-coupling approach in Sec.\ III.

Let us discuss implications of the above general qualitative analysis
to the phase diagram of the extended Hubbard ladder.
From Eqs.\ (\ref{eq:gs's}) and (\ref{eq:mtms}) we write the bare
masses in terms of the coupling constants in the model:
\begin{eqnarray}
m_s 
\!\!&=&\!\!
\frac{1}{2\pi}
\biggl[
    2c_{\rho+} (U-t_\mathrm{pair}+V')
\nonumber\\
&& \quad
 {} + c_{\overline{\rho-}} 
    \left(U-V_\perp+\frac{3}{4}J_\perp
    +t_\mathrm{pair}-4V'\right)
\biggr]
,
\nonumber\\ && 
\label{eq:ms_bare}
\\
m_t 
\!\!&=&\!\!
\frac{1}{2\pi}
\left[
    2 c_{\rho+} \left(V_\perp+\frac{1}{4}J_\perp
                      -\frac{3}{2}V'\right)
\right.\nonumber \\ &&  \quad \left.
  {} + c_{\overline{\rho-}} 
    \left(U-V_\perp+\frac{3}{4}J_\perp
          +t_\mathrm{pair}+2V'\right)
\right]
.
\nonumber\\ &&
\label{eq:mt_bare}
\end{eqnarray}
To simplify the discussion, 
   we assume here that $V_\parallel=V'=t_{\mathrm{pair}}=0$
   and that $\phi_{\rho+}$ is locked at
   $\langle\phi_{\rho+}\rangle=0$ (mod $\pi$), i.e., $c_{\rho+}>0$.
If $U-V_\perp+3J_\perp/4 >0$ ($<0$),
   the phase $\theta_{\rho-}$ is locked at $0$ ($\pi/2$)
   [see Eq.\ (\ref{eq:gc-})] and $c_{\overline{\rho-}}
   =\langle\cos 2\theta_{\rho-}\rangle >0$ $(<0)$.
Thus, the product $c_{\overline{\rho-}} (U-V_\perp+3J_\perp/4)$
   is positive for both positive and negative
   $U-V_\perp+3J_\perp/4$,
   and hence the bare masses $m_s$ and $m_t$ are also positive.
We argue, however, that the Ising criticality is possible 
   due to the mass renormalization effect.
The renormalized mass $\widetilde{m}_s$ can become negative since 
   the coupling constant $g_{\sigma-}$ of the correction term
   in Eq.\ (\ref{eq:ms_tilde}) is 
   given by $g_{\sigma-}=2a(-V_\perp+J_\perp/4)$.
We expect that sufficiently large $V_\perp$ can drive the system
toward the Ising criticality in the $\xi^4$ mode,
even when $t_{\mathrm{pair}}=0$.

In addition to the Ising criticality at large $V_\perp$,
   the Gaussian criticality in the $\theta_{\rho-}$ mode
   should appear at $V_\perp=U+3J_\perp/4$.
Let us find out which phase is realized near the Gaussian
critical line.
When $U-V_\perp+3J_\perp/4=0$,
   the coupling $g_{\sigma-}$ equals $-2U-J_\perp$ and
   the renormalized Ising mass becomes
\begin{equation}
\frac{\widetilde{m}_s}{c_{\rho+}U}=
1
-A\frac{U}{\Lambda}
 \left(1+\frac{3J_\perp}{U}+\frac{2J_\perp^2}{U^2}\right)
 \ln\left(\frac{\Lambda}{U+J_\perp}\right)
,
\end{equation}
where $A$ is a positive constant of order 1.
For small $J_\perp/U$ this renormalized Ising
mass should be positive, and we conclude that the D-Mott
and the S-Mott phases are separated by the Gaussian critical
line (Note that $\widetilde m_t>0$).
As we increase $J_\perp/U$ (or $V_\perp/U$)
   along the Gaussian critical line,
   the negative correction ($\propto g_{\sigma-}$) in
   the mass renormalization increases and eventually $\widetilde m_s$
   can change sign.
Across this Ising transition the D-Mott and S-Mott phases turn
into the SF and CDW phases, respectively.
This implies that a pair of phases surrounding the Gaussian
critical line changes from (D-Mott,S-Mott) to (SF,CDW) at a
tetracritical point as $J_\perp/U$ increases.
This qualitative analysis will be supported in the next subsection
   by a more quantitative renormalization group analysis.

Now we briefly discuss the effect of the 
   pair hopping term $t_\mathrm{pair}$ and next-nearest-neighbor
   repulsion $V'$.
When $V'=0$,
   the Gaussian transition
   takes place at $U-V_\perp+3J_\perp/4+t_{\mathrm{pair}}=0$
   [see Eq.\ (\ref{eq:critical_G})].
Thus for large $t_{\mathrm{pair}}$, we can have a situation
   where $m_s<0$ and $m_t>0$ with
   $U-V_\perp+3J_\perp/4+t_{\mathrm{pair}}\simeq 0$ [see Eqs.\
   (\ref{eq:ms_bare}) and (\ref{eq:mt_bare})], i.e.,
   $t_\mathrm{pair}$ can stabilize the SF state near the Gaussian
   critical line. 
In the case $t_{\mathrm{pair}}=0$, on the other hand,
   we expect that sufficiently large $V'$ can lead to
   a phase with $m_s>0$ and $m_t<0$ i.e., the PDW state,
   if $c_{\rho+}\gg c_{\overline{\rho-}}>0$.

Finally, we discuss implications of our schematic phase diagram
(Fig.\ \ref{fig:diagram}) to the phase diagram of isotropic
   spin-$\frac{1}{2}$ ladder systems, 
   which have been studied intensively in connection with the so-called 
   Haldane's conjecture \cite{Haldane} about the existence of a finite
   energy gap in the integer-spin Heisenberg chain.
By using the abelian bosonization method, it has been shown that four 
   kinds of gapped phases can appear in spin ladder systems
   with various types of exchange interactions.
   \cite{Kim,Gogolin_book}
The possible gapped phases are 
   (1) the rung singlet state, which is known to be realized
   in the isotropic Heisenberg ladder with nearest-neighbor
   antiferromagnetic exchange couplings, 
   (2) the Affleck-Kennedy-Lieb-Tasaki(AKLT)-like spin liquid state,
   in which short-range valence bonds couple spins on neighboring 
   rungs, \cite{AKLT}
   (3) the dimerized state along chain with 
   $\pi$ relative phase, and 
   (4) the dimerized state along chain with zero relative phase. 
Both the rung single state and the AKLT-like state
   are Haldane-type spin liquids with unique ground state
   and no broken local symmetries.
In the dimerized states 
   which are known to be realized when a
   sufficiently strong four-spin interaction is included,
   \cite{Nersesyan1997,Gogolin_book}
   there is spontaneous breaking of the translation ($Z_2$) symmetry
   and the ground state is two-fold degenerate.
In the limit of large $U$ the extended Hubbard ladder
   we analyze in this paper should reduce to a system with
   only the spin degrees of freedom.
This situation corresponds to $g_{\overline{c-}}<0$ 
   [see Eq.\ (\ref{eq:gc-})], i.e., $m_g<0$,
   with $|m_g| \gg |\tilde{m}_s|, |\tilde{m}_t| $.
Under this condition, we still have four phases:
   the SF, D-Mott, PDW, and S'-Mott phases.
From Table \ref{table:phase} (see also 
   Refs.\ \onlinecite{Nersesyan1997,Gogolin_book,Kim}), 
   we can find correspondence
   between the phases in spin ladders and the phases which
   we have obtained in the extended Hubbard ladders:
The rung-singlet and AKLT-like Haldane states correspond to the 
   D-Mott and S'-Mott states, respectively, and
the PDW (SF) state corresponds to the dimerized state along chain with 
   $\pi$ (0) relative phase.
We note that physical pictures of phases in the extended
   Hubbard ladder are consistent with those in spin ladder;
for example, the D-Mott state is nothing bug the rung singlet state,
   as seen in the strong-coupling approach (see Sec.\ III).
The AKLT-like Haldane state, which is known to be realized either
   with plaquette diagonal exchange coupling or with ferromagnetic
   rung exchange,\cite{Kim} would be smoothly 
   connected to the S'-Mott state,
   in which the ground-state wave function consists of
   singlets formed between diagonal sites of plaquettes
   [see Eq.\ (\ref{eq:S'-Mott})] and, moreover,
   has the same topological numbers as the AKLT-like Haldane
   state.\cite{Kim}
The PDW state is nothing but the dimerized state 
   with interchain phase $\pi$ as seen in Fig.\ \ref{fig:diagram},
   which is not a Haldane-type spin liquid since the PDW state
   spontaneously breaks translation symmetry and is two-fold
   degenerate.
In order to discuss phase transitions in spin ladder systems,
   two kinds of string order parameters have been introduced 
   which characterize hidden orders with different topological
   numbers, i.e., the parity of the number of dimers crossing a
   line perpendicular to the two chains. \cite{Nishiyama,Kim}
These string order parameters are different from 
   $\mu_j$ (Eq.\ (\ref{eq:mu_j})), since $\mu_j$ is associated with
   $\exp(i\phi_{\sigma -})$ in the bosonized form 
   while the string order parameters
   introduced in Refs.\ \onlinecite{Kim} and 
   \onlinecite{Nishiyama} are associated with 
   the $\phi_{\sigma +}$ field in our notation.
Since the phase transition associated with the $\phi_{\sigma +}$
   field is related to $\widetilde{m}_t \to 0$,
   we expect that
   the string order parameters introduced 
   in Refs.\ \onlinecite{Kim} and  \onlinecite{Nishiyama} 
   characterize the SU(2)$_2$ criticality or the first-order phase
   transition (double arrows in Fig.\ \ref{fig:diagram}).
In our schematic phase diagram (\ref{fig:diagram})
the phase transition from the rung singlet state to the AKLT Haldane
state can take place (which is actually the case
in the spin-$\frac{1}{2}$ ladder systems\cite{Kim,Hakobyan}),
if the SU(2)$_2$ and the Ising criticalities
appear simultaneously.
This implies that the central charge for the continuous transition
between the rung singlet and the AKLT states is given by
$\frac{3}{2}+\frac{1}{2}=2$.
This transition becomes first order when the marginal interaction
in the triplet Majorana fermion sector is marginally relevant.

\subsection{Renormalization group analysis}

In this subsection, we study the ground-state phase diagram of
the extended Hubbard ladder model using perturbative
RG analysis of the 13 coupling constants appearing
in Eq.\ (\ref{eq:Hint}).
These coupling constants are, however, not independent because
of the 4 constraints coming from the SU(2) symmetry,
Eq.\ (\ref{eq:su2's}).
Accordingly, we have 9 independent RG equations that describe
how the coupling constants scale when we change the lattice constant
$a\to a e^{dl}$.
The 9 independent variables we choose to work with are:
$G_{\rho+}\equiv g_{\rho+}/2\pi v_F$,
$G_{\rho-}\equiv g_{\rho-}/2\pi v_F$,
$G_{\sigma+}\equiv g_{\sigma+}/2\pi v_F$,
$G_{\sigma-}\equiv g_{\sigma-}/2\pi v_F$,
$G_{\alpha}\equiv (g_{c+,s-}-g_{c+,\overline{s-}})/2\pi v_F$,
$G_{\beta}\equiv (g_{\overline{c-},s-}
         -g_{\overline{c-},\overline{s-}})/2\pi v_F$,
$G_A \equiv g_{c+,\overline{c-}}/2\pi v_F$,
$G_B \equiv g_{c+,s+}/2\pi v_F$, and
$G_C \equiv g_{\overline{c-},s+}/2\pi v_F$.
After some algebra we obtain the RG equations:
\begin{eqnarray}    
\frac{d}{dl} G_{\rho+} \!\!&=&\!\! 
 + G_A^2
 + \frac{3}{2} G_B^2
 + \frac{1}{2} G_\alpha^2
,
\label{eq:RGrho+}
\\ 
\frac{d}{dl} G_{\rho-} \!\!&=&\!\! 
 - G_A^2  
 - \frac{3}{2} G_C^2
 - \frac{1}{2} G_\beta^2
,
\\ 
\frac{d}{dl} G_{\sigma+} \!\!&=&\!\! 
 + \frac{1}{2} G_{\sigma+}^2
 + \frac{1}{2} G_{\sigma-}^2
 + G_B^2
 + G_C^2
,
\\ 
\frac{d}{dl} G_{\sigma-} \!\!&=&\!\! 
 + G_{\sigma+} \, G_{\sigma-}
 + G_B \, G_{\alpha}
 + G_C \, G_{\beta}
,
\\
\frac{d}{dl} G_A \!\!&=&\!\! 
  + \frac{1}{2} G_{\rho+} \, G_A
  - \frac{1}{2} G_{\rho-} \, G_A
\nonumber \\ &&
  - \frac{3}{2} G_B \,  G_C
  - \frac{1}{2} G_\alpha G_\beta
, 
\\ 
\frac{d}{dl} G_B \!\!&=&\!\! 
   + \frac{1}{2} G_{\rho+} \, G_B
   + G_{\sigma+} \, G_B
\nonumber \\ &&
   - G_A \, G_C
   + \frac{1}{2} G_{\sigma-} \, G_\alpha 
,
\\ 
\frac{d}{dl} G_C \!\!&=&\!\! 
  - \frac{1}{2} G_{\rho-} \, G_C 
  + G_{\sigma+} \, G_C
\nonumber \\ &&
  - G_A \, G_B
  + \frac{1}{2} G_{\sigma-} \, G_{\beta}
, 
\\
\frac{d}{dl} G_\alpha 
\!\!&=&\!\! 
 + \frac{1}{2} G_{\rho+} \, G_{\alpha}
 + \frac{3}{2} G_B \, G_{\sigma-}
 - G_A \, G_\beta
,
\\
\frac{d}{dl} G_\beta \!\!&=&\!\! 
 - \frac{1}{2} G_{\rho-} \, G_\beta
 + \frac{3}{2} G_C \, G_{\sigma-}
 - G_A \, G_\alpha
.
\qquad
\label{eq:RGbeta}
\end{eqnarray}
These equations are equivalent to the ones reported in
Ref.\ \onlinecite{Lin},
in which another set of 9 independent variables are used:
$b_{11}^\rho = (g_{\rho+} + g_{\rho-})/8$,
$b_{11}^\sigma = - (g_{\sigma+}+g_{\sigma-})/2$,
$b_{12}^\rho = g_\beta/4$,
$b_{12}^\sigma = g_C$,
$f_{12}^\rho = (g_{\rho+}-g_{\rho-})/8$,
$f_{12}^\sigma = -(g_{\sigma+}-g_{\sigma-})$,
$u_{11}^\rho = -g_A/8$,
$u_{12}^\rho = g_\alpha/8$, and
$u_{12}^\sigma = g_B/2$, where $g_\nu=2\pi v_F G_\nu$.

\begin{figure}[t]
\includegraphics[width=7cm]{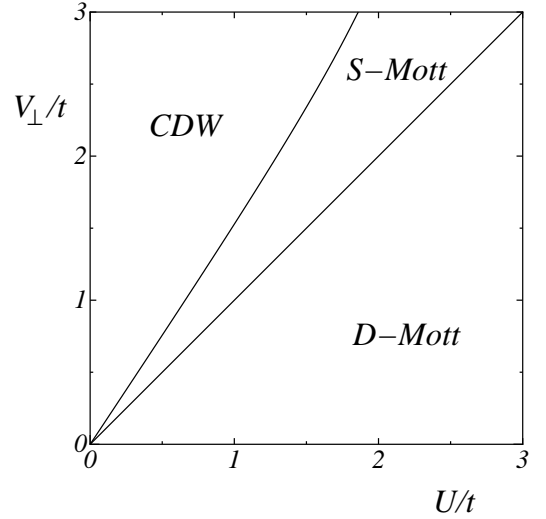}
\caption{
Weak-coupling phase diagram of
$H_{t_\parallel}+H_{t_\perp}+H_\mathrm{int}$ at
$t_\perp=t_\parallel=t$ 
and $J_\perp=0$ obtained from the 1-loop RG equations.
There is a massless mode (C1S0) on the boundary between
 the D-Mott and the S-Mott states  
   while the boundary between the S-Mott and the CDW state
   is C0S$\frac{1}{2}$.
}
\label{fig:phase-b}
\end{figure}

\begin{figure}[t]
\includegraphics[width=7cm]{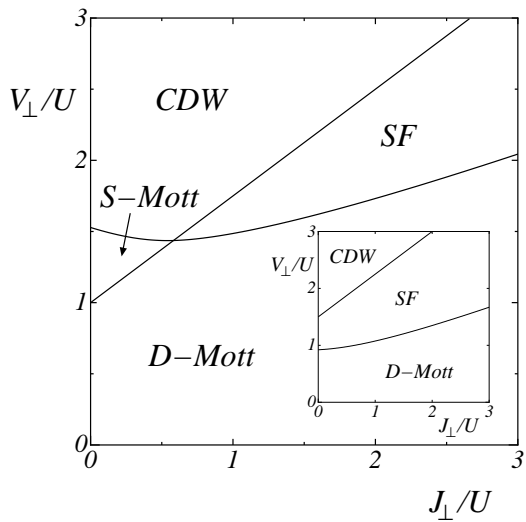}
\caption{
Weak-coupling phase diagram of
$H_{t_\parallel}+H_{t_\perp}+H_\mathrm{int}$ at $t_\perp=t_\parallel$
and $U/t=1$.
This corresponds to Fig.\ \protect{\ref{fig:strong2}}.
Inset shows weak-coupling phase diagram of
$H_{t_\parallel}+H_{t_\perp}+H_\mathrm{int}+H_{\mathrm{pair}}$ 
  at $t_\perp=t_\parallel=t$, $U/t=1$, and $t_{\mathrm{pair}}/t=0.5$.
On the boundaries between the D-Mott and the S-Mott states 
   and between the SF and the CDW states exists a massless mode  C1S0.
A massless mode C0S$\frac{1}{2}$ appears on the boundaries between
the D-Mott and the SF states and between the S-Mott and the CDW states.
The different choice of $U/t$ does not yield
   qualitative changes to this phase diagram.
}
\label{fig:phase-h}
\end{figure}

\begin{figure}[t]
\includegraphics[width=7cm]{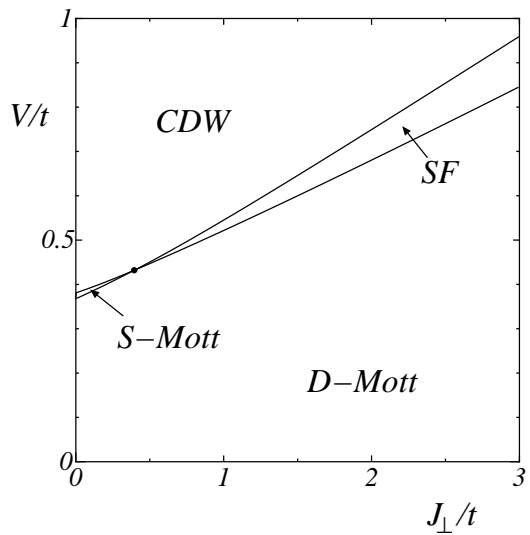}
\caption{
Weak-coupling phase diagram of $H$ for $U/t=1$,
$V_\parallel=V_\perp=V$, and $t_\mathrm{pair}=V'=0$.
The tetracritical point with $C1S\frac{1}{2}$ is at
   $(J_\perp/t,V_\perp/t)\simeq(0.40, 0.43)$.
}
\label{fig:phase-e}
\end{figure}

\begin{figure}[t]
\includegraphics[width=7cm]{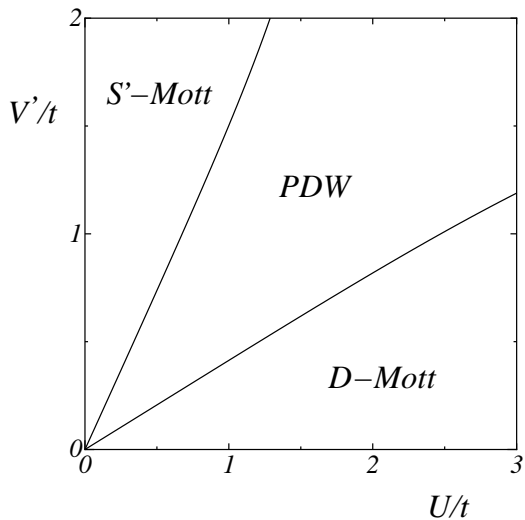}
\caption{
Weak-coupling phase diagram of $H$ on the plane of $U/t$ and 
   $V'/t$ 
   for $V_\parallel=V_\perp=0$, and $J_\perp=t_\mathrm{pair}=0$.
The boundary between the D-Mott state and the PDW state
   is C0S$\frac{3}{2}$, and 
   the boundary between the PDW state and the S'-Mott state
   is C0S$\frac{1}{2}$.
}
\label{fig:phase-c}
\end{figure}

\begin{figure}[t]
\includegraphics[width=7cm]{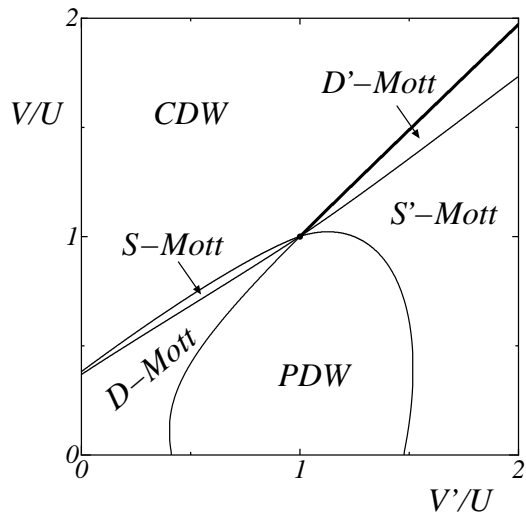}
\caption{
Weak-coupling phase diagram of $H$ on the plane of $V'/U$ and $V/U$ 
   for $U/t=0.5$, $V_\parallel=V_\perp=V$, and
   $J_\perp=t_\mathrm{pair}=0$.
The phase transition between the CDW and S-Mott phases and between the 
   PDW and S'-Mott phases is in the Ising universality class
   (C0S$\frac{1}{2}$).
The phase transition between Mott phases is a Gaussian transition (C1S0).
The boundary between the D-Mott phase and the PDW phase
   is C0S$\frac{3}{2}$ [SU(2)$_2$ criticality].
The transition between the CDW phase and the D'-Mott phase shown by
   the thick solid line is a first-order transition.
}
\label{fig:phase-g}
\end{figure}

Integrating the RG equations (\ref{eq:RGrho+})-(\ref{eq:RGbeta})
   numerically with the initial condition set by the bare coupling
   constants in the extended Hubbard ladder model,
   we find that $G_{\rho+}(l)$ grows most rapidly and becomes
   of order unity first.
At the length scale $l=l_{\rho+}$ where $G_{\rho+}(l_{\rho+})=2$,
   we stop the numerical integration.
Below this energy scale the $\rho+$ mode becomes massive.
We can assume without losing generality 
   that the phase $\phi_{\rho+}$ is locked at 
   $\langle \phi_{\rho+} \rangle=0$ mod $\pi$.
The effective theory at lower energy scale ($l>l_{\rho+}$) is 
   obtained from Eq.\ (\ref{eq:Hint}) through the substitution
   $\cos 2\phi_{\rho+}\to1$,
   $g_{c+,\overline{c-}} \to g_{\overline{c-}}$,
   $g_{c+,s+} \to g_{s+}$,
   $g_{c+,s-} \to g_{s-}$, and
   $g_{c+,\overline{s-}} \to g_{\overline{s-}}$.
We then derive and solve the RG equations for the coupling constants
   in the effective theory to understand the low-energy properties
   of the remaining modes.
The pattern of phase locking can be found from asymptotic low-energy
   behavior of the $g_{\overline{c-}}$, $g_{s+}$,
   $g_{s-}$, and $g_{\overline{s-}}$ in the numerical solution of
   the RG equations.
The phase field $\Phi$ ($=\phi_{\sigma\pm}$ or $\theta_{\rho(\sigma)-}$)
   is locked at $\langle \Phi \rangle =\pi/2$ or 0,
   if the coupling constant $g$
   ($g\in\{g_{\overline{c-}},g_{s+},g_{s-},g_{\overline{s-}}\}$) behaves
   as $g\to + C$ or $-C$ in the low-energy limit, respectively,
   where $C$ is a positive constant of order unity.
Once the configuration of the locked phase fields is determined,
   the resulting ground state is found from
   Table \ref{table:phase-locking}.
The phase diagram of the extended Hubbard ladder obtained in this way
   is shown in Figs.\ \ref{fig:phase-b}--\ref{fig:phase-g}.
We note that this approach reproduces the phase diagram of the SO(5)
   symmetric ladder obtained in earlier studies.
   \cite{Lin,Fjaerestad}
Since the exotic phases like the SF state and the S-Mott state appear
   only for a negative $U$ in this model, we will not further discuss
   it as we concentrate on the case with positive $U$ and $V$ in this
   paper.

Let us first consider the simple case where $U$ and $V_\perp$ are
the only electron-electron interactions.
The phase diagram on the plane of $U/t$ and $V_\perp/t$ is shown 
   in Fig.\  \ref{fig:phase-b}.
In this and other phase diagrams shown below, all the modes are gapped
everywhere except on the phase boundaries.
With the standard notation C$n$S$m$ of representing a state having
$n$ massless charge modes and $m$ massless spin
modes,\cite{Balents1996} the three phases in Fig.~\ref{fig:phase-b}
are characterized as the ``C0S0'' phase.\cite{Balents1996,Lin}
The phase boundary between the D-Mott state and the S-Mott state
   is the U(1) Gaussian critical line of the $\rho-$ mode (C1S0),
   which is given by $V_\perp=U$; see
   Eq.\ (\ref{eq:critical_G}) with $J_\perp=0$.
The phase boundary between the S-Mott state and the CDW state
   is the Ising critical line of the spin $\sigma-$ mode,
   which is C0S$\frac{1}{2}$.
This weak-coupling phase diagram is similar to 
   Fig.\ \ref{fig:strong1} 
   obtained from the strong-coupling approach.

Next, we include the AF exchange coupling $J_\perp$.
The phase diagram on the plane of $J_\perp/U$ and $V_\perp/U$
   at $U/t=1$ is shown in Fig.\ \ref{fig:phase-h}.
A different choice of $U/t$ does not lead to qualitative changes
   in the $J_\perp/U$ vs $V_\perp/U$ phase diagram.
An interesting new feature is that the SF phase shows up
between the D-Mott phase and the CDW phase.
This is in agreement with the qualitative analysis of
   the previous subsection, where it is found that
   the exchange interaction $J_\perp$ suppresses
   the S-Mott phase and helps the SF phase appear.
The Gaussian criticality of the $\rho-$ mode (C1S0) emerges on
   the almost straight phase boundary between the D-Mott phase and
   the S-Mott phase and between the SF phase and the CDW phase.
This critical line is given by $V_\perp/U=1+3J_\perp/4U$,
   in accordance with Eq.\ (\ref{eq:critical_G}).
The phase boundary between the D-Mott phase and the SF phase and
between the S-Mott phase and the CDW phase
   is the Ising criticality C0S$\frac{1}{2}$.
A tetracritical point of C1S$\frac{1}{2}$ appears at the point
where the two kinds of phase boundaries cross.
The inset of Fig.\ \ref{fig:phase-h} shows the phase diagram at
$t_\mathrm{pair}=0.5t$.
We see clearly that the pair-hopping favors the SF phase over the
S-Mott phase.
In the strong-coupling perturbation theory,
we have introduced the pair-hopping term $H_{\mathrm{pair}}$
to stabilize the SF state.
This is not necessary, however, in the weak-coupling approach,
where the pair-hopping process is effectively generated from
the second-order process in the rung hopping $t_\perp$.
In fact, we can show that positive pair-hopping terms are generated
   in the renormalization-group procedure in the SF
   phase.\cite{Tsuchiizu2001}

Next we turn on the nearest-neighbor Coulomb repulsion in the
   leg direction, $V_\parallel$.
The phase diagram for $V_\parallel=V_\perp(\equiv V)$ is shown
   in Fig.\ \ref{fig:phase-e}.
Even though the additional $V_\parallel$ interaction strongly favors
the CDW state, a small region of the S-Mott phase still remains in
between the D-Mott phase and the CDW phase.
Besides this quantitative modification the phase diagram is not changed
qualitatively, and, in particular, the critical properties at the phase
boundaries are the same as in Figs.\ \ref{fig:phase-b} and
\ref{fig:phase-h}.
Using the density matrix renormalization group method,
Vojta \textit{et al.}\cite{Vojta1999} determined the phase boundary
between the CDW state and a state with homogeneous charge density for
the model we used for Fig.\ \ref{fig:phase-e}.
At $U=1.5t$ they observed a transition to the CDW state around
$U/V\approx2.9$, which is not very different from the
phase boundary at $J_\perp=0$ in Fig.\ \ref{fig:phase-e}.
The transition is, however, found to be first order for $U\ge4t$ in
their numerical results, which is different from the continuous
transition we found in the weak-coupling analysis.
A possible source of this discrepancy might be the neglect of
irrelevant operators with canonical dimension 4 that could become
important for strong couplings as in the single chain
case.\cite{Tsuchiizu2002} 

Finally, we include next-nearest-neighbor Coulomb
   repulsion $V'$, Eq.\ (\ref{eq:HVprime}).
Figures \ref{fig:phase-c} and \ref{fig:phase-g} show
   the $V'$-$U$ and $V$-$V'$ phase diagrams.
In agreement with the discussion in the previous subsection,
   the PDW phase appears as $V'$ is increased.
At even larger $V'$ the S'-Mott phase and the D'-Mott phase appear
   in Figs.\ \ref{fig:phase-c} and \ref{fig:phase-g}.
On the phase boundary between the D-Mott state and the
   PDW state appears the SU(2)$_2$ criticality;
we have confirmed in our numerical calculation that the coupling
   $g_{\sigma+}$ in Eq.\ (\ref{eq:Heff_spin}) is negative, i.e.,
   marginally irrelevant.
We have thus established that the two-particle interaction $V'$ can
drive the system to the SU(2)$_2$ criticality.

Figure \ref{fig:phase-g} shows a rich phase diagram containing the
four Mott phases and the two density-wave phases.
We note that in Fig.\ \ref{fig:phase-g} the six phase boundaries
   meet at $V=V'=U$, which corresponds to C2S2.
This happened because, within our approximation, all the coupling
constants in Eq.\ (\ref{eq:Hint}) except $g_{\rho+}$ vanish
when $U=V=V'$,
$t_\perp=t_\parallel$, and $J_\perp=t_\mathrm{pair}=0$.
If $t_\perp\ne t_\parallel$, or if higher-order contributions to the
$g$'s are included,\cite{Tsuchiizu2002} this special situation might
not occur.
In Fig.\ \ref{fig:phase-g} the phase boundaries between the Mott
phases are C1S0 (Gaussian criticality), while the CDW--S-Mott
and PDW--S'-Mott phase boundaries are C0S$\frac12$ (Ising criticality).
The phase boundary between the PDW phase and the D-Mott phase is
C0S$\frac32$ [SU(2)$_2$ criticality] as in Fig.\ \ref{fig:phase-c}.
Finally, the phase transition between the CDW phase and the D'-Mott
phase is found to be first order; we have confirmed that the coupling
$g_{\sigma+}$ in Eq.\ (\ref{eq:Heff_spin}) is positive and marginally
relevant.
Even though Fig.\ \ref{fig:phase-g} is obtained from the weak-coupling
RG equations, we think that the phase diagram is reliable since we
have confirmed that the $V/U$-$V'/U$ phase diagram is not changed much 
when $U/t$ is varied.

\section{Conclusions}\label{sec:summary}

In this paper 
   we have studied the half-filled generalized
   Hubbard ladder with the inter-site
   Coulomb repulsion and the exchange interaction
   by using the strong-coupling perturbation theory and
   the weak-coupling bosonization method.
In the strong-coupling approach the SF state is described as an AF
   ordered state of the Ising model where pseudo-spins
   represent the currents flowing along the rungs.
We have shown that the SF state can appear next to
   the CDW state and the D-Mott state in the phase diagram and that
the quantum phase transition between the SF state and the D-Mott state
   is in the Ising universality class.
We have also established the Ising transition between the S-Mott and
the CDW phases and the Gaussian transition between the D-Mott and the
S-Mott phases.
In the weak-coupling approach we have shown that in general the model
can accommodate total of eight insulating phases at half-filling,
four density-wave phases and four Mott phases (Fig.\ \ref{fig:diagram}).
The universality class of the phase transitions among these phases is
determined.
In particular, we have shown that the SU(2)$_2$ criticality
   with the central charge $c=3/2$ is induced by
   the next-nearest-neighbor Coulomb repulsion $V'$, which drives
the system from the D-Mott phase to the PDW phase (Figs.\
   \ref{fig:phase-c} and \ref{fig:phase-g}).
When $V'$ is further increased, the S'-Mott phase and the D'-Mott
   phase, which correspond to the quantum disordered states of the PDW
   phase and the FDW phase, show up (Fig.\ \ref{fig:phase-c}).

When this manuscript was almost completed, we became aware of the
work by Wu \textit{et al.},\cite{Fradkin2002} where the 8 insulating
phases in Sec.\ IV are obtained independently.

\acknowledgments

We thank M.\ Sigrist, C.\ Mudry, and H.\ Tsunetsugu for
   helpful discussions.
We also thank E.\ Orignac for pointing out to us the importance of the
   marginal operator in the analysis of the SU(2)$_2$ criticality.
One of the authors (AF) thanks S.\ Chakravarty and M.\ Troyer for
   enlightening discussions at the Aspen Center for Physics.
This work was supported in part by Grant-in-Aid for
   Scientific Research on Priority Areas (A) from The Ministry of
   Education, Culture, Sports, Science and Technology (No.\ 12046238).


\begin{thebibliography}{}
\bibitem{transitions}
It is well known that 
   the universality classes of critical properties in quantum 1D
   systems are classified by the conformal field theory
   (CFT).
For example,
   the free boson theory (the Gaussian model) is a conformal theory 
   with the central charge $c=1$,  while
   the free (real) fermion theory, which is known to describe
   the Ising criticality, has the central charge $c=1/2$.
   \cite{Gogolin_book}
In the ladder model we consider, the system can have massless
   excitations on the phase transition boundaries.
Thus the critical properties of the various quantum phase transitions
   are classified in terms of the CFT.
The possible transition types in our model are
  the $c=1$ Gaussian criticality in the charge sector, and
  the $c=1/2$ Ising and $c=3/2$ SU(2)$_2$ 
  criticalities in the spin sector.
The SU(2)$_2$ criticality is described by a $k=2$ SU(2)
   Wess-Zumino-Witten model and is equivalent to 3 massless Majorana
   fermions.
The critical exponents of these critical theories are known and can be 
   found in the literature.
\end{thebibliography}
\end{document}